\documentclass[twocolumn, aps,prl, floatfix, superscriptaddress]{revtex4-1}
\usepackage{graphicx}

\usepackage{color}\usepackage{enumerate}
\usepackage{amsmath}\usepackage{amssymb}\usepackage{stmaryrd}
\usepackage{multirow}
\usepackage{hhline}
\usepackage{tikz}
\usetikzlibrary{matrix}

\usepackage{tcolorbox, accents}
\usepackage{tikz-cd}
\usetikzlibrary{cd}
\usepackage{longtable,booktabs}
\usepackage[colorlinks=true,citecolor=cyan,linkcolor=magenta]{hyperref}
\usepackage{url}
\usepackage{color}
\usepackage{enumerate} 
\usepackage{amsmath}\usepackage{amssymb}\usepackage{stmaryrd}
\usepackage{multirow}
\usepackage{float}
\usepackage{mathrsfs}
\usepackage{bbold}
\usepackage[toc,page]{appendix}
\usepackage{longtable,booktabs}
 \usepackage{epsfig}
\usepackage{graphicx} 
\usepackage{feynmf}
\usepackage{psfrag} 
\usepackage{slashed}
\usepackage{mathtools}
\usepackage[normalem]{ulem}
\usepackage{amsmath,amssymb} 
\usepackage{colordvi} 
\usepackage{hyperref}
\usepackage{setspace}
\usepackage[all]{hypcap}
\usepackage{natbib}
\usepackage{subfigure}

\hypersetup{
   bookmarks=true,         
   unicode=false,          
 pdftoolbar=true,        
    pdfmenubar=true,        
    pdffitwindow=false,     
    pdfstartview={FitH},    
    pdftitle={My title},    
    pdfauthor={Author},     
    pdfsubject={Subject},   
    pdfcreator={Creator},   
    pdfproducer={Producer}, 
    pdfkeywords={keyword1} {key2} {key3}, 
    pdfnewwindow=true,      
    colorlinks=true,       
    linkcolor=cyan,          
    citecolor=purple,        
    filecolor=magenta,      
    urlcolor=magenta           
}
\usepackage{natbib} 
\newcommand\SP[1]{{\color{blue} #1}}
\def\be{\begin{equation}}
\def\bea{\begin{eqnarray}}
\def\eea{\end{eqnarray}}
\def\ee{\end{equation}}

\newcommand{\ms}{\mathsf}

\newcommand{\imagi}{{i}}

\begin{document}

\title{`Unhinging' the surfaces of higher-order topological insulators and superconductors}

\author{Apoorv Tiwari}
\affiliation{Department of Physics, University of Zurich, Winterthurerstrasse 190, 8057 Zurich, Switzerland}
\affiliation{Condensed Matter Theory Group, Paul Scherrer Institute, CH-5232 Villigen PSI, Switzerland}

\author{Ming-Hao Li}
\affiliation{Department of Physics, ETH Zurich, 8093 Zurich, Switzerland}

\author{B.A. Bernevig}
\affiliation{Department of Physics, Princeton University, Princeton, New Jersey 08540, USA}

\author{Titus Neupert}
\affiliation{Department of Physics, University of Zurich, Winterthurerstrasse 190, 8057 Zurich, Switzerland}

\author{S. A. Parameswaran}
\affiliation{Rudolf Peierls Centre for Theoretical Physics, Clarendon Laboratory, University of Oxford, Oxford, OX1 3PU, UK}

\begin{abstract}
We show that the chiral Dirac and Majorana hinge modes in three-dimensional higher-order topological insulators (HOTIs) and superconductors (HOTSCs)  can be gapped while preserving the protecting $\ms{C}_{2n}\mathcal T$ symmetry upon the introduction of non-Abelian surface topological order. In both cases, the topological order on a single side surface breaks time reversal symmetry, but appears with its time-reversal conjugate on alternating sides in a $\ms{C}_{2n}\mathcal T$ preserving pattern. In the absence of the HOTI/HOTSC bulk, such a pattern necessarily involves gapless chiral modes on hinges between $\ms{C}_{2n}\mathcal T$-conjugate domains. However, using a combination of $K$-matrix and anyon condensation arguments, we show that on the boundary of a 3D HOTI/HOTSC these topological orders are fully gapped and hence `anomalous'. Our results suggest that new patterns of surface and hinge states can be engineered by selectively introducing topological order only on specific surfaces.
\end{abstract}
\maketitle
\emph{Introduction.---} 
A defining aspect of topological phases of matter is the bulk-boundary correspondence. This predicts the existence of gapless excitations on the boundary of an insulating phase from the bulk electronic structure alone, irrespective of boundary details.
Initially, it was believed that the correspondence inevitably requires  
gapless  surface excitations as long as system and boundary both respect the protecting
symmetries of the  
bulk topological phase.
 This would imply, for example, that a three-dimensional (3D) electronic topological insulator (TI), protected by time-reversal ($\mathcal{T}$) and $\mathsf{U}(1)$ charge conservation symmetry always hosts a surface Dirac fermion if $\mathcal{T}\ltimes \mathsf{U}(1)$ is respected. However, there is another possibility~\cite{Vish_2013, Burnell_2014, Metlitski_2013, PhysRevB.92.125111, Bonderson_2013, Wang_2013, Chen_2014, Maissam_2013, Chen_2015}:  the  3D TI  surface can be fully gapped with $\mathcal{T}\ltimes \mathsf{U}(1)$ symmetry intact, if it hosts a 
  topologically ordered state 
 \cite{wen2004quantum, Wen_TO_95, Kitaev_Toric_Code, Levin_Wen, Kitaev_anyons, Kitaev_TEE}, i.e., an intrinsically interacting phase with emergent fractionalized excitations. Thus, the complete bulk-boundary correspondence for a 3D TI states that a symmetry-preserving surface \emph{either} carries a gapless Dirac fermion \emph{or} the appropriate surface topological order (STO). Both  these surface terminations cancel the bulk {anomaly} arising from the $\mathbf{E}\cdot\mathbf{B}$ electromagnetic response, although only the former has been experimentally observed. As a corollary, 
  the STO  cannot be realized with the same symmetries in a  purely 2D system. 
This generalized bulk-boundary correspondence also applies to other 3D  topological phases such as $\mathcal{T}$-symmetric
topological superconductors (TSCs)~\cite{Fidkowski_2013, Wang_Senthil_2014, Max_2014, Maissam_2016, Wang_2017, Tachikawa_2017a, Tachikawa_2017, Meng_2018}. 


A different type of  
 bulk-boundary correspondence emerges in higher-order topological insulators and superconductors (HOTIs/HOTSCs)~\cite{Benalcazar61, Wlad_2017, Schindlereaat0346, PhysRevLett.119.246402, Brouwer_et_al_2017, parameswaran2017a, Khalaf_2018, Brouwer_2018, Brouwer_2019, schindler2018higher, imhof2018topolectrical, Yizhi_2018, Rasmussen_Lu_2018, 2019arXiv190107579G}. These 
  bulk-gapped phases of matter carry topologically protected boundary modes on corners or hinges, instead of surfaces (in 3D).  
  Such protection requires a spatial symmetry  that maps between patches of the surface, making the interplay of topology and crystal symmetry~\cite{PhysRevX.7.011020,PhysRevB.92.081304, PhysRevB.95.125107, PhysRevB.96.205106,PhysRevX.8.011040, 2019arXiv190106195K} central to the study of HOTIs/HOTSCs.
 
In this Letter, we 
generalize the higher-order bulk-boundary correspondence to include the possibility of STO. 
Specifically, we study 3D topological insulators and superconductors with chiral hinge modes --- the HOTI/HOTSC analogs of integer quantum Hall states or $\ms{p}+i\ms{p}$ superconductors. For concreteness, we consider cases where 
the protecting symmetry is $\ms{C}_{2n}\mathcal T$
, i.e., the product of a $(2n)$-fold rotation and time-reversal $\mathcal{T}$. {In other words,  $\mathcal{T}$ and $\ms{C}_{2n}$ are individually broken but their product remains unbroken.} (Here $n$ is a positive integer, and $n\leq 3$ for any 3D space group). Nontrivial HOTI/HOTSC phases with these symmetries support chiral fermionic modes 
on each of $2n$ hinges in a $\ms{C}_{2n}$-symmetric geometry with open boundary conditions in the rotation plane. 
Such phases have a $\mathbb{Z}_2$ topological classification:
while a single chiral fermionic mode is stable and symmetry-protected in the non-interacting limit, {\it two} chiral Dirac/Majorana modes on each hinge can be gapped out  by pasting copies of the integer quantum Hall phase with $\nu=\pm 1$ (for the HOTI) or $\ms{p}\pm i\ms{p}$ 2D topological superconductors (for the HOTSC) in alternating fashion on the surfaces while preserving $\ms{C}_{2n}\mathcal T$ symmetry. It is natural to ask: can these modes be gapped while preserving symmetry in an interacting system?



We answer this question in the affirmative by 
 constructing symmetry-preserving STOs that `unhinge' the gapless modes on the HOTI/HOTSC surfaces. In the HOTI case, we 
  leverage the $K$-matrix formulation of coupled Luttinger liquids to show that the hinge is gapped. For the HOTSC we cannot use this method, but instead 
  map the question  to an auxiliary anyon condensation problem.
  We close with a discussion of why the resulting $\ms{C}_{2n}\mathcal T$ STOs we construct are {\it anomalous} ---  in that they can be fully gapped only on the surface of a HOTI/HOTSC ---
 and identify  directions for future work.

\emph{Higher order TI.} --- We begin by constructing a symmetry-preserving STO for the $\ms{C}_{2n}\mathcal T$ HOTI. Since we are discussing insulators, in addition to $\ms{C}_{2n}\mathcal T$ we must impose $\ms{U}(1)$ charge conservation symmetry (implicit in the noninteracting classification~\cite{Schindlereaat0346}), otherwise the hinge could be simply gapped by depositing $p\pm ip$ superconductors on alternating surfaces. 
Each fermionic hinge mode carries $\ms{U}(1)$ electric charge $q=+1$ (in units of the electron charge $e$) and has chiral central charge $c_-=1$~\cite{francesco2012conformal, 1991Ginsparg}. {These respectively quantify the chiral hinge transport of charge and heat}. In order to respect $\ms{C}_{2n}\mathcal T$ symmetry, we must impose an alternating pattern of topological order $\mathcal{A}$ and its $\mathcal{T}$-conjugate $\bar{\mathcal{A}}$ on adjacent side surfaces; however, the STO on the top/bottom surface (that we denote $\mathcal{A}_{\mathcal{T}}$) should preserve $\ms{C}_{2n}\mathcal T$. In order for the side STOs to cancel the contribution of the hinge, the Hall conductance $\sigma_{xy}^{\mathcal{A}}=-\sigma_{xy}^{\bar{\mathcal{A}}}=\frac12$ in units of $e^2/h$ and the chiral central charge $c^{\mathcal{A}}_{-}=-c^{\bar{\mathcal{A}}}_{-}=\frac12$. Thus, $\mathcal{A}, \bar{\mathcal{A}}$ must be chiral and non-Abelian. The same constraints emerge when constructing STO for TIs~\cite{Bonderson_2013, Chen_2014}, where a close cousin of the Pfaffian topological order~\cite{Fradkin_98, Fendley_2007, Bishara_2008}  known as the $\mathcal T$-Pfaffian was constructed. Notably, as it has $c_-\neq 0$ 
 the $\mathcal T$-Pfaffian necessarily breaks $\mathcal{T}$ when realized in a purely 2D system, but it can preserve $\mathcal{T}$ on the 2D surface of a 3D TI~\cite{Chen_2014}. 

\begin{figure}[bt]
\centering
\includegraphics[width=\columnwidth]{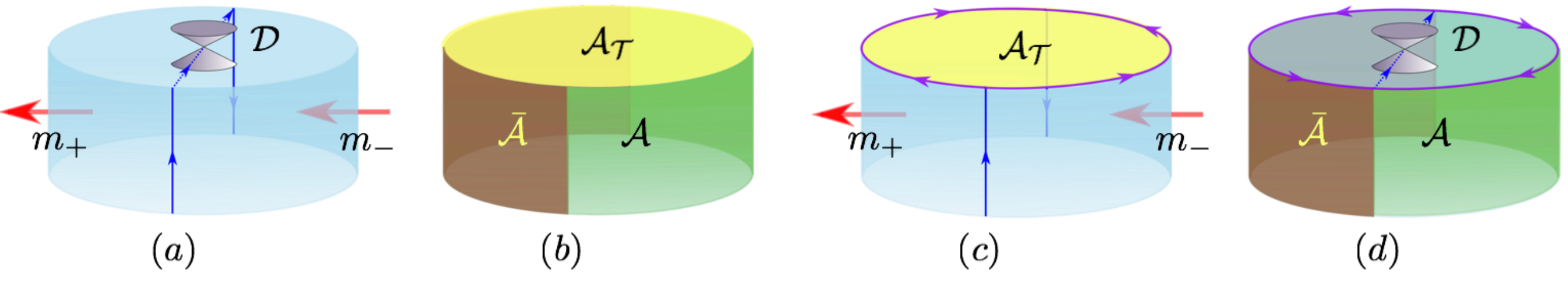}
\caption{Possible surface terminations of $\ms{C}_2\mathcal{T}$ HOTI/HOTSC. (a) The underlying free-fermion phase has $\mathcal{T}$-breaking surface gaps $m_{\pm}$ in a $\ms{C}_2\mathcal{T}$ pattern, leading to chiral modes on side hinges and a 2D Dirac/Majorana fermion $\mathcal{D}$ stablized by $\ms{C}_2\mathcal{T}$  on the gapless top surface. (b)  Non-Abelian STO $\mathcal{A}_{\mathcal{T}}$ only on the top surface leads to a `wire frame' of chiral modes on all hinges. (c) Adding $\mathcal{T}$-breaking  2D analogs $\mathcal{A}, \bar{\mathcal{A}}$ of $\mathcal{A}_{\mathcal{T}}$ on  $\ms{C}_2\mathcal{T}$-related sides fully gaps the boundary. (d) Non-Abelian STO only on the sides yields chiral edge modes co-existing with a 2D Dirac/Majorana fermion on the top surfaces. (Bottom surfaces follow a similar pattern, omitted for clarity).}
\label{fig:gapcases}
\end{figure}

A fully gapped surface termination for the HOTI can be constructed by taking the top/bottom STO $\mathcal{A}_{\mathcal{T}}$ to be the $\mathcal{T}$-Pfaffian, and the side STO $\mathcal{A}$ to be the 2D $\mathcal{T}$-breaking phase with chiral edge modes that has the same anyon content as the $\mathcal{T}$-Pfaffian, and $\bar{\mathcal{A}}$ the $\mathcal{T}$-conjugate of $\mathcal{A}$. To motivate this choice, we note that the free-fermion $\ms{C}_{2n}\mathcal T$ HOTI emerges upon introducing $\mathcal{T}$-breaking gaps (denoted $m_{\pm}$ where the sign indicates that of the $\mathcal{T}$-breaking) on alternating surfaces of a first-order TI in a $\ms{C}_{2n}\mathcal T$-preserving manner (Fig.~\ref{fig:gapcases}a depicts a $\ms{C}_{2}\mathcal T$ example). The top/bottom surfaces then each host a single 2D Dirac fermion. 
By imposing $\mathcal{A}_\mathcal{T}$ on the top/bottom surfaces we gap out the surface Dirac fermion while preserving $\ms{C}_{2n}\mathcal T$; however, this introduces modes with   
$|c_-|= |q| =1/2$  
on the top and bottom hinges between $\mathcal{A}_\mathcal{T}$ and $m_{\mp}$, that combine with the side hinges in a `wire frame' pattern (Fig.~\ref{fig:gapcases}b).  The edges between the $\mathcal{T}$-Pfaffian and the time-reversal-breaking region $m_{\pm}$ are respectively identical to those between its 2D analogues $\mathcal{A}, \bar{\mathcal{A}}$ and vacuum~\cite{Bonderson_2013}. Accordingly, we may gap the top and bottom hinges by adding  $\mathcal{A},  \bar{\mathcal{A}}$ to the $m_{-}$ and $m_+$ surfaces respectively, as this yields the necessary pattern of counterpropagating modes. Finally, the boundary between $\mathcal{A}$, $\bar{\mathcal{A}}$ oriented as in Fig.~\ref{fig:gapcases}c carries $c_- = q =-1$, which cancels the side hinges. [We can shrink gapless top/bottom regions to a set of 1D chiral modes that slice across them, while preserving $\ms{C}_{2n}\mathcal T$. For $n=1$ this leaves one chiral mode that encircles the sample, and the analysis is just that for the side hinge. For $n>1$ the surface chiral mode pattern is more complicated. Introducing $\mathcal{A}_\mathcal{T}$ makes our approach  $n$- independent.]

Before explicitly verifying the hinge gapping, we review some properties of the $\mathcal{T}$-Pfaffian and its 2D $\mathcal{T}$-breaking analogues. These all have  identical bulk anyon content: a subset of the product of topological quantum field theories (TQFTs) $ \ms{U}(1)_8 \times \overline{\text{Ising}}$ with anyon types $1_{j
}, \psi_{j}$ (with $j=0,2,4,6$) and $\sigma_{j}$ (with $j=1,3,5,7$), and braiding and fusion rules derived from the direct product theory~\cite{SupMat}.   This is a spin TQFT~\cite{Rubenstein_2016, Bhardwaj2017, aasen2017} containing a  charge $1$  `transparent' fermion, $\psi_{4}$ that braids trivially with all other particles. In conventional TQFTs, such particles are identified with vacuum, but this is precluded here as $\psi_4$ is a fermion;  instead it is identified with the physical electron. The vacuum of a spin TQFT 
is `graded' by fermion parity, 
meaning that  only those anyons in $\ms{U}(1)_8\times \overline{\text{Ising}}$ that braid trivially with $\psi_{4}$ are retained 
  (see Tab.~\ref{tab:Tpfaf}).
A TQFT with these anyons  is necessarily chiral and 
 can be realized in a $\mathcal{T}$-preserving manner only on the surface of a 3D TI, where it is termed the  $\mathcal{T}$-Pfaffian (our choice of $\mathcal{A}_{\mathcal{T}}$). {On the 3D TI surface, $\mathcal{T}$ interchanges 
$1_{2}\leftrightarrow \psi_2$ and $1_6\leftrightarrow \psi_6$,
 and squares to $-1$ on $\psi_4$;
all other anyons are $\mathcal{T}$-invariant~\footnote{$\mathcal{A}_{\mathcal{T}}$ is the `$\mathcal{T}$-Pfaffian$_+$', the STO of  the free-fermion TI. A distinct  `$\mathcal{T}$-Pfaffian$_-$'   with sign-reversed topological spins and $\mathcal{T}^2$ actions (where defined) for  $1_{2,6},\psi_{2,6}$ and $\sigma_j$ yields an STO for an intrinsically interacting HOTI.}}. While  $\mathcal{A}_{\mathcal{T}}$ cannot have an edge with vacuum, 
it has a chiral edge with $\mathcal{T}$-breaking regions $m_{\pm}$ on the TI surface.
 {$\mathcal{T}$-{\it breaking} TQFTs  with  identical anyon content {\it can} be realized in 2D with chiral edges to vacuum: these are the 2D analogues $\mathcal{A},\bar{\mathcal{A}}$ of the $\mathcal{T}$-Pfaffian.
  The edges  all share the same  Lagrangian~\cite{Bonderson_2013}
\begin{align}
\mathcal{L}^{\mathsf{a}}_\pm =\frac{2}{4\pi}\partial_{x}\phi^{\mathsf{a}}(\partial_t\mp v\partial_x) \phi^{\mathsf{a}}+ \imagi\psi^{\mathsf{a}}(\partial_t\pm v'\partial_x)\psi^{\mathsf{a}},
\label{eq:T_pfaf_edge}
\end{align}   
consisting of a chiral $\ms{U}(1)$ boson  $\phi^{\mathsf{a}}$ and a counterpropagating chiral Majorana fermion  $\psi^{\mathsf{a}}$, where $\pm$ denotes the sign of both $c_-$ and $q$. {(We adopt a Lagrangian description to conveniently describe chiral modes.)} We label edge fields  between $\mathcal{A}$, $\bar{\mathcal{A}}$ and vacuum by ${\mathsf{a}} = {\mathsf A}, \bar{\mathsf A}$, and those between the $\mathcal{T}$-breaking side surfaces $m_{\pm}$  
and $\mathcal{A}_{\mathcal{T}}$ by  ${\mathsf{a}}= {\mathsf m}_{\pm}$. Additionally we enforce a $\mathbb{Z}^{\mathsf{a}}_2$ gauge symmetry $\psi^{\mathsf{a}} \mapsto  - \psi^{\mathsf{a}}, \phi^{\mathsf{a}}\mapsto \phi^{\mathsf{a}} +\frac{\pi}{2}$, which identify  $\psi^{\mathsf{a}} e^{-2i\phi^{\mathsf{a}}}$ as the edge electron operator  ~\cite{Meng_2018, SupMat}.
\begin{table}[]
\begin{center}
\small
\begin{tabular}{c|c|c|c|c|c|c|c|c|c|c|c|c}
             $a\rightarrow$          & $1_0$ & $\psi_0$ & ${1_2}$    & $\psi_2$ & $1_4$ & $\psi_4$ & $1_6$    & $\psi_6$ & $\sigma_1$ & $\sigma_3$ & $\sigma_5$ & $\sigma_7$ \\ \hhline{-|-|-|-|-|-|-|-|-|-|-|-|-}
$e^{i\theta_a}$        & $1$   & $-1$     & $-i$     & $i$      & $1$   & $-1$     & $-i$     & $i$      & $1$        & $-1$       & $-1$       & $1$        \\ \hline
$Q_a$ & $0$   & $0$      & $1/2$    & $1/2$    & $1$   & $1$      & $3/2$   & $3/2$   & $1/4$      & $3/4$     & $5/4$     & $7/4$    \\ \hline
$\mathcal{T}(a)$          & $1_0$ & $\psi_0$ & $\psi_2$ & $1_2$    & $1_4$ & $\psi_4$ & $\psi_6$ & $1_6$    & $\sigma_1$ & $\sigma_3$ & $\sigma_5$ & $\sigma_7$ \\ \hline
$\mathcal{T}_a^2$        & $1$   & $1$      &          &          & $-1$  & $-1$     &          &          & $1$        & $-1$       & $-1$       & $1$\\ 
\end{tabular}
\label{tab:Tpfaf}
\end{center}
\caption{Anyons  $a$  in the $\mathcal{T}$-Pffaffian and 2D analogues and their topological spin $e^{i\theta_a}$, $\mathsf{U}(1)$ charge $Q_a$ (units of $e$),  time-reversal partner $\mathcal{T}_a$ and `Kramers sign' $\mathcal{T}_a^2$ (where applicable).}
\end{table}
Any  top/bottom hinge is a `composite' of the edges between  $\mathcal{A}$ (or $\bar{\mathcal{A}}$) and vacuum, and between $\mathcal{A}_{\mathcal{T}}$ and $m_{-}$   (or $m_+$), and is hence  described by $\mathcal{L}^{\mathsf A}_s+ \mathcal{L}^{\mathsf m_-}_{-s}$ (or $\mathcal{L}^{\bar{\mathsf A}}_s + \mathcal{L}^{\mathsf m_+}_{-s}$), with $s=\pm$. The two theories in each sum are mutually  $\mathcal{T}$-conjugate (i.e., acting with $\mathcal{T}$ on one  yields the other), so $c_-= q =0$, and can be gapped without breaking $\mathsf{U}(1)$  symmetry. 
At each side hinge, the bulk HOTI contributes a  chiral mode 
\begin{equation}
\mathcal{L}_{\pm}^h = \frac{1}{4\pi}\partial_{x}\varphi (\partial_t\mp u\partial_x) \varphi.\label{eq:hingechiral}
\end{equation}
We next observe that the effective Lagrangian at a single side hinge (see Fig.~\ref{fig:C2n}a) that includes the chiral modes from both the HOTI bulk and from $\mathcal{A}, \mathcal{\bar{A}}$ takes the form $\mathcal{L} = \mathcal{L}_{-}^{\mathsf A} + \mathcal{L}_{+}^{\bar{\mathsf A}} + \mathcal{L}_-^h$. Since $\mathsf{A},\bar{\mathsf{A}}$ are $\mathcal{T}$-conjugates, $\mathcal{L}_{-}^{\mathsf A} + \mathcal{L}_{+}^{\bar{\mathsf A}}$ is really just two copies of $\mathcal{L}_{-}^{\mathsf A}$.  The two Majorana modes therefore co-propagate with each other and with the hinge mode $\varphi$, but counterpropagate relative to the chiral boson fields $\phi^{\mathsf A}, \phi^{\bar{\mathsf A}}$. Therefore, we may combine $\psi^{\mathsf A}, \psi^{\bar{\mathsf A}}$ into a single chiral Dirac fermion, that we then bosonize into a compact chiral neutral boson via $e^{i\phi}\sim \psi^{\ms{a}}+ i\psi^{\bar{\ms{a}}}$. This series of manipulations 
{recasts} the edge as a coupled Luttinger-liquid theory~\cite{wen2004quantum, Wen_TO_95} described by the $K$-matrix $K=\text{diag}(1, -2, -2, 1)$ in the boson basis $\Phi:=(\phi, \phi^{\ms{A}},\phi^{\bar{\ms{A}}}, \varphi)^{\ms{T}}$, where the coefficients follow from Eqs.~\eqref{eq:T_pfaf_edge} and \eqref{eq:hingechiral}. The $\ms{U}(1)$ electric charges of the boson fields are captured by the vector $q=(0,1,1,1)^{\ms{T}}$. The combined theory has vanishing Hall conductance  $\sigma_{xy}=q^{\ms{T}}K^{-1}q=0$, and 
the chiral central charge $c_- = \text{signature}(K) =0$, meaning there is no 
immediate obstruction (i.e., due to Hall or thermal Hall responses) to gapping the hinge theory  $\mathcal{L}$. 
{We do so by adding}
 $\Delta \mathcal{L} =\sum_{i} \lambda_i \cos\left[\ell_i^{\ms{T}}\Phi +\alpha_i\right]$   and driving all the $\lambda_i$ to strong coupling
 ~\cite{Haldane_1995, Levin_2013, Maissam_2013, Juven_2015}. 
The combination of fields $\ell_i^{\ms{T}}\Phi$  must (i)~correspond to bosonic non-chiral edge operators  which is true if $\ell_i^{\ms{T}}K^{-1}\ell_i =0$; (ii)   be non-fractional i.e $\ell_{i}\in K \mathbb Z^{4}$; (iii) be charge neutral so that the gapped phase preserves $\ms{U}(1)$, requiring $\ell_i^{\ms{T}}K^{-1}q=0$.  Finally the ${\mathbb Z}^{\mathsf{A}}_{2}\times {\mathbb Z}^{\bar{\mathsf{A}}}_{2} $ gauge symmetry must also be satisfied. 
First, we condense $\ell_1 = (0,4,4,4)^{\ms{T}}$,
 {This locks} %
  the two independent gauge transformations to act together as $\phi\mapsto \phi+\pi$, $\phi^{\ms A(\bar{\ms{A}})}\mapsto \phi^{\ms A(\bar{\ms{A}})}\pm\pi/2$~\cite{SupMat}. 
This lets us condense $\ell_2 = (2,2,-2,0)^{\ms{T}}$ which is invariant under this unbroken subgroup of $\mathbb Z_{2}^{\ms{A}}\times \mathbb Z_{2}^{\bar{\ms{A}}}$. Since $\ell_{1,2}$ satisfy all the above criteria and $\ell^{\ms{T}}_1 K^{-1}\ell_2 =0$,  they can simultaneously flow to strong coupling, leading to a symmetric, gapped, non-degenerate edge.

 \emph{Higher order TSC.} --- We now consider the $\ms{C}_{2n}\mathcal{T}$-symmetric HOTSC, that hosts an alternating pattern of $c_-=\frac{1}{2}$ Majorana hinge modes. 
In analogy with the HOTI, to construct an STO we  should start with the `parent' first-order topological phase, namely the $\nu=1$ class DIII TSC, whose surface hosts a single Majorana cone in the free-fermion limit. However, the STO for this phase is complicated~\cite{Meng_2018}. A simpler route is to recognize that only the {\it parity} of $\nu$ is relevant to the $\ms{C}_{2n}\mathcal{T}$-HOTSC: we can 
change  hinge chiral central charge in multiples of $1/2$ by gluing $\ms{p}\pm i\ms{p}$ superconductors to alternating side surface in a $\ms{C}_{2n}\mathcal{T}$-preserving manner (i.e. it  suffices that $c_-^{\mathcal{A}} = -c_-^{\bar{\mathcal{A}}} = \frac14 \mod \frac12$). Since a pure surface perturbation changes $\nu\rightarrow \nu+2$,  
 we can instead consider a related $\ms{C}_{2n}\mathcal{T}$-HOTSC obtained by  decorating the $\nu=3$ DIII first-order TSC with $\mathcal{T}$-breaking  domains $\ms{m}_\pm$ on side surfaces,  yielding a chiral hinge  mode with three Majoranas ($|c_-|=3/2$). The $\nu=3$ STO in class DIII is the $\ms{SO}(3)_6$ TQFT, which may be viewed as the integer spin sector of the $\mathsf{SU}(2)_6$ theory~\cite{Fidkowski_2013,Wang_2017}. Similar reasoning as in the HOTI case suggests that we should take this as the topological order $\mathcal{A}_{\mathcal{T}}$ for the top/bottom surfaces, and then pattern its 2D $\mathcal{T}$-breaking analogues $\mathcal{A}$ and $\bar{\mathcal{A}}$ in a $\ms{C}_{2n}\mathcal{T}$-preserving fashion on the side-surface.  It will be convenient to also glue three copies of $\ms{p}+i\ms{p}$ superconductors in a $\ms{C}_{2n}\mathcal{T}$-preserving pattern on the side surfaces. We now show that the side hinge is gapped; then, by `Kirchoff's law' for edge modes, we can infer that the top/bottom hinges are gapped.
A single  $\ms{SO}(3)_6$ edge is described by a chiral Wess-Zumino-Witten theory with $c_-=9/4$, so the side hinge is more complicated and unlike the HOTI case cannot be  rewritten in terms of chiral bosons. Therefore,  we cannot use the $K$-matrix approach and need some other strategy to proceed. One route is via `conformal embedding'~\cite{Meng_2018}. Here we instead use anyon condensation to infer the edge structure.
 
We first  impose periodic boundary conditions along  the ${\ms C}_{2n}$ axis, to focus only on the alternating pattern of side STOs $\mathcal{A}, \bar{\mathcal{A}}$.  The question of gappability now reduces to (i) determining the hinge mode between 
\SP{$\mathcal{T}$}-conjugate topological orders $\mathcal{A}, \bar{\mathcal{A}}$, and (ii) showing that it can gap the hinge modes contributed by the combination of the bulk HOTI and the  3 additional  $\ms{p}\pm i\ms{p}$ states decorating the side surfaces. Step (i) may be further simplified by  `folding' $\bar{\mathcal{A}}$ across the hinge which maps the boundary between ${\mathcal{A}}$ and  $\bar{\mathcal{A}}$ to an edge between $\mathcal{A}\times \mathcal{A}$ and the vacuum  (see Fig.~\ref{fig:C2n}b). We can infer the minimal edge theory by condensing a maximal  subset of anyons in the bulk of the folded theory $\mathcal{A}\times \mathcal{A}$.

 \begin{figure}[bt]
\centering
\includegraphics[width=0.9\columnwidth]{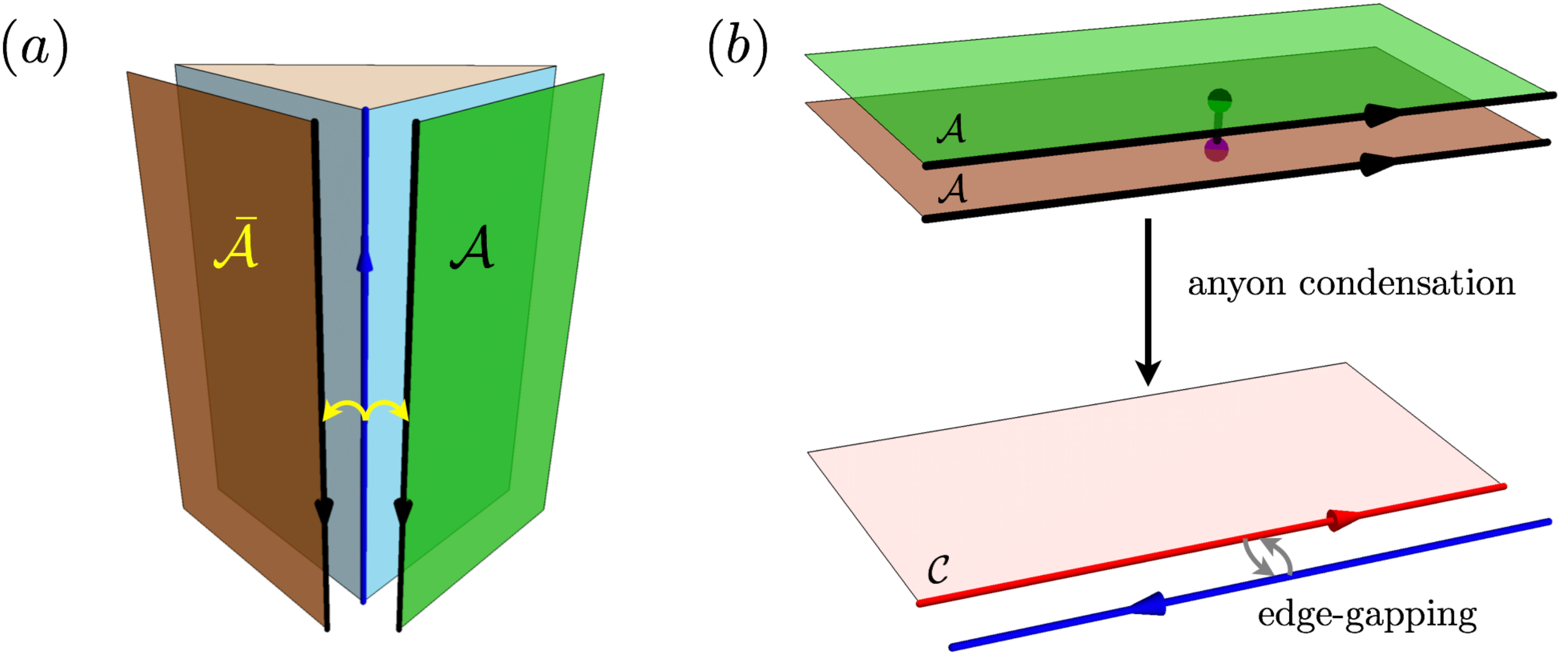}
\caption{Constructing symmetry-preserving gapped side surfaces of a $\ms{C}_{2n}\mathcal{T}$-HOTI/HOTSC. (a)  2D STO $\mathcal{A}$ and its {$\mathcal{T}$}-conjugate $\bar{\mathcal{A}}$ are imposed on alternate surfaces, so that their edge modes combine with the hinge mode yield a fully gapped boundary. $\mathcal{A}$, $\bar{\mathcal{A}}$ are 2D analogues of the $\mathcal{T}$-symmetric 
STO $\mathcal{A}_{\mathcal{T}}$ 
for the `parent' first-order HOTI/HOTSC. (b) By `folding' across the hinge, this can also be viewed as the process of condensing some subset of anyons in $\mathcal{A}\times \mathcal{A}$ to yield a chiral  topological phase with no bulk anyons ($\mathcal{C}$) whose edge mode then gaps out the hinge mode of  the bulk HOTI/HOTSC.}
\label{fig:C2n}
\end{figure}

We  first validate this approach for the HOTI. 
We denote anyons in $\mathcal A\times \mathcal A$ by elements in the set $\{1_{j}^{\ms{A}},\psi_{j}^{\ms{A}},\sigma_{j}^{\ms{A}}\} \times \{1_{j}^{\bar{\ms{A}}},\psi_{j}^{\bar{\ms{A}}},\sigma_{j}^{\bar{\ms{A}}}\}$ (see Tab.~\ref{tab:Tpfaf}; we label anyons in the second copy of $\mathcal{A}$ by $\bar{\ms A}$, to indicate their origin in $\bar{\mathcal{A}}$ before folding). Following \cite{Chen_2014}, we perform a two-step condensation procedure. First, we condense the bosons $
\{1_{2}^{\ms{A}}\psi_{6}^{\bar{\ms{A}}}, 
\psi_{2}^{\ms{A}}1_{6}^{\bar{\ms{A}}},
\psi_{6}^{\ms{A}}1_{2}^{\bar{\ms{A}}},
1_{6}^{\ms{A}}\psi_{2}^{\bar{\ms{A}}}, 
1_{4}^{\ms{A}}1_{4}^{\bar{\ms{A}}},
\psi_{0}^{\ms{A}}\psi_{0}^{\bar{\ms{A}}},
\psi_{4}^{\ms{A}}\psi_{4}^{\bar{\ms{A}}}
\}$. %
This confines all sectors in  $\mathcal A\times \mathcal A$ whose topological spin is not a good quantum number, leaving only
the Abelian anyons 
$\{1_{0}^{\ms{A}}1_{0}^{\bar{\ms{A}}}, \psi_{0}^{\ms{A}}1_{0}^{\bar{\ms{A}}}, \psi_{4}^{\ms{A}}1_{0}^{\bar{\ms{A}}}, 1_{4}^{\ms{A}}1_{0}^{\bar{\ms{A}}}\}\simeq \{1_{0}^{\ms{A}}, \psi_{0}^{\ms{A}}, \psi_{4}^{\ms{A}}, 1_{4}^{\ms{A}}\}$ and the non-Abelian anyons $\sigma_{1}^{\ms{A}}\sigma_{3}^{\bar{\ms{A}}}$ and $\sigma_{1}^{\ms{A}}\sigma_{7}^{\bar{\ms{A}}}$. Crucially, the non-Abelian anyon sectors split into two Abelian anyons each in the condensed theory. Therefore the condensed theory contains eight Abelian anyons, four of which are charge neutral while the remaining four carry charge $+1$~\cite{SupMat}. The neutral anyons correspond to the toric code topological order \cite{Kitaev_Toric_Code}. The charged anyons correspond to a copy of the toric code obtained from the neutral anyons by fusing with the physical electron $\psi_4^{\ms{A}}$. Next, we condense the `$e$'-particle in the charge-neutral copy of the toric code. This gaps out the entire theory except for $\{1_0^{\ms{A}},\psi_4^{\ms{A}}\}$. The surviving sectors correspond to a bulk theory whose edge has a single chiral fermionic mode with unit $\ms{U}(1)$ charge (since $c_-=1$ is unchanged by condensation). We then use this to gap the counter-propagating hinge mode of the bulk HOTI~\cite{SupMat}. Note that no additional surface decorations were needed in this case.

We now turn to the HOTSC case where $\mathcal{A}$ corresponds to the $\ms{SO}(3)_6$ TQFT, which contains four anyons labeled $j=\left\{0,1,2,3\right\}$ with topological spin $\left\{+1,+i,-i,-1\right\}$ respectively. {The surface of the 3D class DIII TSC, admits a time-reversal symmetric realization of $\ms{SO}(3)_6$ 
wherein $\mathcal{T}$ exchanges the anyons $j=1$ and $j=2$, leaves $j=0$ invariant, and squares to $-1$ on $j=3$, which is identified with  the physical electron.} 
As in the HOTI case we label the anyons in the folded theory $\mathcal{A}\times \mathcal{A}$ (equivalent to  operators on the hinge/domain wall between $\mathcal A$ and $\bar{\mathcal A}$) by $(j^{\ms{A}},j^{\bar{\ms{A}}}) \in \left\{0,1,2,3\right\} \times \left\{0,1,2,3\right\}$.  $\mathcal{A}\times \mathcal{A}$ contains four  mutually local bosons with labels $\left\{(00),(33),(21),(12)\right\}$. Condensing these four bosons confines all  remaining anyons except for $\left\{(03), (30), (11), (22)\right\}$. In the condensed theory these are all fermions and may be identified with a single fermionic sector, which we denote $\mathfrak f$. 
{We can verify}~\cite{SupMat} that $\mathfrak f$ is neutral and local i.e braids trivially with itself. The domain wall between $\mathcal A$ and $\bar{\mathcal A}$ thus reduces to a local neutral fermion with 
$c_{-}= \frac{9}{2}$ (recall condensation preserves $c_-$). We combine this with the 9 non-interacting  Majorana modes (3+3  from $\ms{p}\pm i\ms{p}$ SCs   decorating adjacent side surfaces, and 3 from the $\nu=3$ HOTSC bulk) to fully gap the side hinge.


 \emph{Discussion.---}  {We have constructed fully-gapped  $\ms{C}_{2n}\mathcal T$-preserving STOs for  HOTI/HOTSCs, exemplifying the generalized higher-order bulk-boundary correspondence. The STOs are anomalous and cannot be realized in strictly 2D. For instance, imposing STO only on the top surface (Fig.~\ref{fig:gapcases}b) yields a chiral mode pattern that is impossible on any orientable 2D manifold, but is consistent on a HOTI surface because of the hinges. Similarly, if we consider the  $\ms{C}_{2n}\mathcal T$-preserving alternating pattern of $\mathcal{T}$-breaking orders on the side surfaces only (with, e.g.~periodic boundary conditions along $z$), 
 we see that in 2D these would host gapless modes at every hinge, but these are canceled by those from the bulk when the same pattern is realized on the 3D HOTI/HOTSC side surface. This also gives us insight into the $\ms{C}_{2n}\mathcal T$-preserving gapless surface state present on the top/bottom surfaces of the HOTI: by gapping only the side surfaces with STOs, 
 we see that the top/bottom surfaces host 
 a chiral Dirac/Majorana in their 2D bulk, but also have a characteristic  $\ms{C}_{2n}\mathcal T$-preserving pattern of edge modes (Fig.~\ref{fig:gapcases}d); this warrants further study. Junction structures  --- e.g., the `wire frame'  where imposing STO only on the top/bottom surfaces yields a symmetric  `beam splitter' dividing a non-interacting chiral mode into two intrinsically 
 interacting ones --- are natural with the lower symmetry of HOTIs/HOTSCs, offering a promising line of investigation.}
 
{Although so far most predicted HOTIs/HOTSCs are weakly interacting, they likely have  a rich set of interacting counterparts similar to the topological Kondo and Mott insulators proposed in the first-order case. For example,  a natural way to break $\mathcal{T}$ while preserving $\ms{C}_{2n}\mathcal T$ is to trigger surface magnetic order, which requires interactions. Our results are likely relevant to experiments in the strongly-correlated regime where interactions can gap out the hinge modes, leaving only the more subtle signatures of higher-order topology described here. Furthermore, our ideas  generalize to analogous higher order symmetry-protected topological phases (HOSPTs) 
in bosonic/spin systems that lack a `free' limit. For instance,  perturbing the bosonic class DIII TSC~\cite{Vish_2013}  with time-reversal breaking in a $\ms{C}_{2n}\mathcal T$-preserving manner yields a bosonic $\ms{C}_{2n}\mathcal T$- HOSPT. 
The relevant STO is obtained by taking $\mathcal{A}_{\mathcal{T}}$ to be the ``3-fermion $\mathbb{Z}_2$'' state~\cite{Burnell_2014} that cancels the bulk anomaly of the first-order DIII TSC and $\mathcal{A}, \bar{\mathcal{A}}$ its $\mathcal{T}$-breaking 2D analogues. Extensions 
 to second-order SPTs protected by inversion~\cite{TLNBP-unpub} and to third-order 3D SPTs with gapless corner modes, are avenues for future work.}

  \emph{Acknowledgements.} --- We are grateful to D.~Aasen, M.~Barkeshli, L.~Fidkowski,   F.~Pollmann, A.C.~Potter, A.~Nahum, S.~Ramamurthy and S.H.~Simon for useful discussions, and are especially grateful to M. Cheng for very useful discussions and correspondence on Ref.~\cite{Meng_2018}. SAP and TN thank the Max Planck Institute for the Structure and Dynamics of Matter for hospitality during the initiation of this work.  We acknowledge support from the 
   European Research Council (ERC) under the European Union Horizon 2020 Research and Innovation Programme [Grant Agreements Nos.~ 757867-PARATOP (TN) and ~804213-TMCS (SAP) and the Marie Sklodowska Curie Grant Agree-
386 ment No. 701647 (A. T.)]. BAB acknowledges support from the
Department of Energy de-sc0016239, Simons Investigator
Award, the Packard Foundation, the Schmidt Fund for
Innovative Research, NSF EAGER grant DMR-1643312,
ONR-N00014-14-1-0330, and NSF-MRSEC DMR-1420541.
 
\bibliography{HOTI_interaction}

\begin{thebibliography}{69}%
\makeatletter
\providecommand \@ifxundefined [1]{%
 \@ifx{#1\undefined}
}%
\providecommand \@ifnum [1]{%
 \ifnum #1\expandafter \@firstoftwo
 \else \expandafter \@secondoftwo
 \fi
}%
\providecommand \@ifx [1]{%
 \ifx #1\expandafter \@firstoftwo
 \else \expandafter \@secondoftwo
 \fi
}%
\providecommand \natexlab [1]{#1}%
\providecommand \enquote  [1]{``#1''}%
\providecommand \bibnamefont  [1]{#1}%
\providecommand \bibfnamefont [1]{#1}%
\providecommand \citenamefont [1]{#1}%
\providecommand \href@noop [0]{\@secondoftwo}%
\providecommand \href [0]{\begingroup \@sanitize@url \@href}%
\providecommand \@href[1]{\@@startlink{#1}\@@href}%
\providecommand \@@href[1]{\endgroup#1\@@endlink}%
\providecommand \@sanitize@url [0]{\catcode `\\12\catcode `\$12\catcode
  `\&12\catcode `\#12\catcode `\^12\catcode `\_12\catcode `\%12\relax}%
\providecommand \@@startlink[1]{}%
\providecommand \@@endlink[0]{}%
\providecommand \url  [0]{\begingroup\@sanitize@url \@url }%
\providecommand \@url [1]{\endgroup\@href {#1}{\urlprefix }}%
\providecommand \urlprefix  [0]{URL }%
\providecommand \Eprint [0]{\href }%
\providecommand \doibase [0]{http://dx.doi.org/}%
\providecommand \selectlanguage [0]{\@gobble}%
\providecommand \bibinfo  [0]{\@secondoftwo}%
\providecommand \bibfield  [0]{\@secondoftwo}%
\providecommand \translation [1]{[#1]}%
\providecommand \BibitemOpen [0]{}%
\providecommand \bibitemStop [0]{}%
\providecommand \bibitemNoStop [0]{.\EOS\space}%
\providecommand \EOS [0]{\spacefactor3000\relax}%
\providecommand \BibitemShut  [1]{\csname bibitem#1\endcsname}%
\let\auto@bib@innerbib\@empty
\bibitem [{\citenamefont {Vishwanath}\ and\ \citenamefont
  {Senthil}(2013)}]{Vish_2013}%
  \BibitemOpen
  \bibfield  {author} {\bibinfo {author} {\bibfnamefont {A.}~\bibnamefont
  {Vishwanath}}\ and\ \bibinfo {author} {\bibfnamefont {T.}~\bibnamefont
  {Senthil}},\ }\href {\doibase 10.1103/PhysRevX.3.011016} {\bibfield
  {journal} {\bibinfo  {journal} {Phys. Rev. X}\ }\textbf {\bibinfo {volume}
  {3}},\ \bibinfo {pages} {011016} (\bibinfo {year} {2013})}\BibitemShut
  {NoStop}%
\bibitem [{\citenamefont {Burnell}\ \emph {et~al.}(2014)\citenamefont
  {Burnell}, \citenamefont {Chen}, \citenamefont {Fidkowski},\ and\
  \citenamefont {Vishwanath}}]{Burnell_2014}%
  \BibitemOpen
  \bibfield  {author} {\bibinfo {author} {\bibfnamefont {F.~J.}\ \bibnamefont
  {Burnell}}, \bibinfo {author} {\bibfnamefont {X.}~\bibnamefont {Chen}},
  \bibinfo {author} {\bibfnamefont {L.}~\bibnamefont {Fidkowski}}, \ and\
  \bibinfo {author} {\bibfnamefont {A.}~\bibnamefont {Vishwanath}},\ }\href
  {\doibase 10.1103/PhysRevB.90.245122} {\bibfield  {journal} {\bibinfo
  {journal} {Phys. Rev. B}\ }\textbf {\bibinfo {volume} {90}},\ \bibinfo
  {pages} {245122} (\bibinfo {year} {2014})}\BibitemShut {NoStop}%
\bibitem [{\citenamefont {Metlitski}\ \emph {et~al.}(2013)\citenamefont
  {Metlitski}, \citenamefont {Kane},\ and\ \citenamefont
  {Fisher}}]{Metlitski_2013}%
  \BibitemOpen
  \bibfield  {author} {\bibinfo {author} {\bibfnamefont {M.~A.}\ \bibnamefont
  {Metlitski}}, \bibinfo {author} {\bibfnamefont {C.~L.}\ \bibnamefont {Kane}},
  \ and\ \bibinfo {author} {\bibfnamefont {M.~P.~A.}\ \bibnamefont {Fisher}},\
  }\href {\doibase 10.1103/PhysRevB.88.035131} {\bibfield  {journal} {\bibinfo
  {journal} {Phys. Rev. B}\ }\textbf {\bibinfo {volume} {88}},\ \bibinfo
  {pages} {035131} (\bibinfo {year} {2013})}\BibitemShut {NoStop}%
\bibitem [{\citenamefont {Metlitski}\ \emph {et~al.}(2015)\citenamefont
  {Metlitski}, \citenamefont {Kane},\ and\ \citenamefont
  {Fisher}}]{PhysRevB.92.125111}%
  \BibitemOpen
  \bibfield  {author} {\bibinfo {author} {\bibfnamefont {M.~A.}\ \bibnamefont
  {Metlitski}}, \bibinfo {author} {\bibfnamefont {C.~L.}\ \bibnamefont {Kane}},
  \ and\ \bibinfo {author} {\bibfnamefont {M.~P.~A.}\ \bibnamefont {Fisher}},\
  }\href {\doibase 10.1103/PhysRevB.92.125111} {\bibfield  {journal} {\bibinfo
  {journal} {Phys. Rev. B}\ }\textbf {\bibinfo {volume} {92}},\ \bibinfo
  {pages} {125111} (\bibinfo {year} {2015})}\BibitemShut {NoStop}%
\bibitem [{\citenamefont {Bonderson}\ \emph {et~al.}(2013)\citenamefont
  {Bonderson}, \citenamefont {Nayak},\ and\ \citenamefont
  {Qi}}]{Bonderson_2013}%
  \BibitemOpen
  \bibfield  {author} {\bibinfo {author} {\bibfnamefont {P.}~\bibnamefont
  {Bonderson}}, \bibinfo {author} {\bibfnamefont {C.}~\bibnamefont {Nayak}}, \
  and\ \bibinfo {author} {\bibfnamefont {X.-L.}\ \bibnamefont {Qi}},\ }\href
  {\doibase 10.1088/1742-5468/2013/09/p09016} {\bibfield  {journal} {\bibinfo
  {journal} {Journal of Statistical Mechanics: Theory and Experiment}\ }\textbf
  {\bibinfo {volume} {2013}},\ \bibinfo {pages} {P09016} (\bibinfo {year}
  {2013})}\BibitemShut {NoStop}%
\bibitem [{\citenamefont {Wang}\ \emph {et~al.}(2013)\citenamefont {Wang},
  \citenamefont {Potter},\ and\ \citenamefont {Senthil}}]{Wang_2013}%
  \BibitemOpen
  \bibfield  {author} {\bibinfo {author} {\bibfnamefont {C.}~\bibnamefont
  {Wang}}, \bibinfo {author} {\bibfnamefont {A.~C.}\ \bibnamefont {Potter}}, \
  and\ \bibinfo {author} {\bibfnamefont {T.}~\bibnamefont {Senthil}},\ }\href
  {\doibase 10.1103/PhysRevB.88.115137} {\bibfield  {journal} {\bibinfo
  {journal} {Phys. Rev. B}\ }\textbf {\bibinfo {volume} {88}},\ \bibinfo
  {pages} {115137} (\bibinfo {year} {2013})}\BibitemShut {NoStop}%
\bibitem [{\citenamefont {Chen}\ \emph {et~al.}(2014)\citenamefont {Chen},
  \citenamefont {Fidkowski},\ and\ \citenamefont {Vishwanath}}]{Chen_2014}%
  \BibitemOpen
  \bibfield  {author} {\bibinfo {author} {\bibfnamefont {X.}~\bibnamefont
  {Chen}}, \bibinfo {author} {\bibfnamefont {L.}~\bibnamefont {Fidkowski}}, \
  and\ \bibinfo {author} {\bibfnamefont {A.}~\bibnamefont {Vishwanath}},\
  }\href {\doibase 10.1103/PhysRevB.89.165132} {\bibfield  {journal} {\bibinfo
  {journal} {Phys. Rev. B}\ }\textbf {\bibinfo {volume} {89}},\ \bibinfo
  {pages} {165132} (\bibinfo {year} {2014})}\BibitemShut {NoStop}%
\bibitem [{\citenamefont {Barkeshli}\ \emph {et~al.}(2013)\citenamefont
  {Barkeshli}, \citenamefont {Jian},\ and\ \citenamefont {Qi}}]{Maissam_2013}%
  \BibitemOpen
  \bibfield  {author} {\bibinfo {author} {\bibfnamefont {M.}~\bibnamefont
  {Barkeshli}}, \bibinfo {author} {\bibfnamefont {C.-M.}\ \bibnamefont {Jian}},
  \ and\ \bibinfo {author} {\bibfnamefont {X.-L.}\ \bibnamefont {Qi}},\ }\href
  {\doibase 10.1103/PhysRevB.88.235103} {\bibfield  {journal} {\bibinfo
  {journal} {Phys. Rev. B}\ }\textbf {\bibinfo {volume} {88}},\ \bibinfo
  {pages} {235103} (\bibinfo {year} {2013})}\BibitemShut {NoStop}%
\bibitem [{\citenamefont {Chen}\ \emph {et~al.}(2015)\citenamefont {Chen},
  \citenamefont {Burnell}, \citenamefont {Vishwanath},\ and\ \citenamefont
  {Fidkowski}}]{Chen_2015}%
  \BibitemOpen
  \bibfield  {author} {\bibinfo {author} {\bibfnamefont {X.}~\bibnamefont
  {Chen}}, \bibinfo {author} {\bibfnamefont {F.~J.}\ \bibnamefont {Burnell}},
  \bibinfo {author} {\bibfnamefont {A.}~\bibnamefont {Vishwanath}}, \ and\
  \bibinfo {author} {\bibfnamefont {L.}~\bibnamefont {Fidkowski}},\ }\href
  {\doibase 10.1103/PhysRevX.5.041013} {\bibfield  {journal} {\bibinfo
  {journal} {Phys. Rev. X}\ }\textbf {\bibinfo {volume} {5}},\ \bibinfo {pages}
  {041013} (\bibinfo {year} {2015})}\BibitemShut {NoStop}%
\bibitem [{\citenamefont {Wen}(2004)}]{wen2004quantum}%
  \BibitemOpen
  \bibfield  {author} {\bibinfo {author} {\bibfnamefont {X.-G.}\ \bibnamefont
  {Wen}},\ }\href@noop {} {\emph {\bibinfo {title} {Quantum field theory of
  many-body systems: from the origin of sound to an origin of light and
  electrons}}}\ (\bibinfo  {publisher} {Oxford University Press on Demand},\
  \bibinfo {year} {2004})\BibitemShut {NoStop}%
\bibitem [{\citenamefont {Wen}(1995)}]{Wen_TO_95}%
  \BibitemOpen
  \bibfield  {author} {\bibinfo {author} {\bibfnamefont {X.-G.}\ \bibnamefont
  {Wen}},\ }\href {\doibase 10.1080/00018739500101566} {\bibfield  {journal}
  {\bibinfo  {journal} {Advances in Physics}\ }\textbf {\bibinfo {volume}
  {44}},\ \bibinfo {pages} {405} (\bibinfo {year} {1995})},\ \Eprint
  {http://arxiv.org/abs/https://doi.org/10.1080/00018739500101566}
  {https://doi.org/10.1080/00018739500101566} \BibitemShut {NoStop}%
\bibitem [{\citenamefont {Kitaev}(2003)}]{Kitaev_Toric_Code}%
  \BibitemOpen
  \bibfield  {author} {\bibinfo {author} {\bibfnamefont {A.}~\bibnamefont
  {Kitaev}},\ }\href {\doibase https://doi.org/10.1016/S0003-4916(02)00018-0}
  {\bibfield  {journal} {\bibinfo  {journal} {Annals of Physics}\ }\textbf
  {\bibinfo {volume} {303}},\ \bibinfo {pages} {2 } (\bibinfo {year}
  {2003})}\BibitemShut {NoStop}%
\bibitem [{\citenamefont {Levin}\ and\ \citenamefont {Wen}(2005)}]{Levin_Wen}%
  \BibitemOpen
  \bibfield  {author} {\bibinfo {author} {\bibfnamefont {M.~A.}\ \bibnamefont
  {Levin}}\ and\ \bibinfo {author} {\bibfnamefont {X.-G.}\ \bibnamefont
  {Wen}},\ }\href {\doibase 10.1103/PhysRevB.71.045110} {\bibfield  {journal}
  {\bibinfo  {journal} {Phys. Rev. B}\ }\textbf {\bibinfo {volume} {71}},\
  \bibinfo {pages} {045110} (\bibinfo {year} {2005})}\BibitemShut {NoStop}%
\bibitem [{\citenamefont {Kitaev}(2006)}]{Kitaev_anyons}%
  \BibitemOpen
  \bibfield  {author} {\bibinfo {author} {\bibfnamefont {A.}~\bibnamefont
  {Kitaev}},\ }\href {\doibase https://doi.org/10.1016/j.aop.2005.10.005}
  {\bibfield  {journal} {\bibinfo  {journal} {Annals of Physics}\ }\textbf
  {\bibinfo {volume} {321}},\ \bibinfo {pages} {2 } (\bibinfo {year} {2006})},\
  \bibinfo {note} {january Special Issue}\BibitemShut {NoStop}%
\bibitem [{\citenamefont {Kitaev}\ and\ \citenamefont
  {Preskill}(2006)}]{Kitaev_TEE}%
  \BibitemOpen
  \bibfield  {author} {\bibinfo {author} {\bibfnamefont {A.}~\bibnamefont
  {Kitaev}}\ and\ \bibinfo {author} {\bibfnamefont {J.}~\bibnamefont
  {Preskill}},\ }\href {\doibase 10.1103/PhysRevLett.96.110404} {\bibfield
  {journal} {\bibinfo  {journal} {Phys. Rev. Lett.}\ }\textbf {\bibinfo
  {volume} {96}},\ \bibinfo {pages} {110404} (\bibinfo {year}
  {2006})}\BibitemShut {NoStop}%
\bibitem [{\citenamefont {Fidkowski}\ \emph {et~al.}(2013)\citenamefont
  {Fidkowski}, \citenamefont {Chen},\ and\ \citenamefont
  {Vishwanath}}]{Fidkowski_2013}%
  \BibitemOpen
  \bibfield  {author} {\bibinfo {author} {\bibfnamefont {L.}~\bibnamefont
  {Fidkowski}}, \bibinfo {author} {\bibfnamefont {X.}~\bibnamefont {Chen}}, \
  and\ \bibinfo {author} {\bibfnamefont {A.}~\bibnamefont {Vishwanath}},\
  }\href {\doibase 10.1103/PhysRevX.3.041016} {\bibfield  {journal} {\bibinfo
  {journal} {Phys. Rev. X}\ }\textbf {\bibinfo {volume} {3}},\ \bibinfo {pages}
  {041016} (\bibinfo {year} {2013})}\BibitemShut {NoStop}%
\bibitem [{\citenamefont {Wang}\ and\ \citenamefont
  {Senthil}(2014)}]{Wang_Senthil_2014}%
  \BibitemOpen
  \bibfield  {author} {\bibinfo {author} {\bibfnamefont {C.}~\bibnamefont
  {Wang}}\ and\ \bibinfo {author} {\bibfnamefont {T.}~\bibnamefont {Senthil}},\
  }\href {\doibase 10.1103/PhysRevB.89.195124} {\bibfield  {journal} {\bibinfo
  {journal} {Phys. Rev. B}\ }\textbf {\bibinfo {volume} {89}},\ \bibinfo
  {pages} {195124} (\bibinfo {year} {2014})}\BibitemShut {NoStop}%
\bibitem [{\citenamefont {{Metlitski}}\ \emph {et~al.}(2014)\citenamefont
  {{Metlitski}}, \citenamefont {{Fidkowski}}, \citenamefont {{Chen}},\ and\
  \citenamefont {{Vishwanath}}}]{Max_2014}%
  \BibitemOpen
  \bibfield  {author} {\bibinfo {author} {\bibfnamefont {M.~A.}\ \bibnamefont
  {{Metlitski}}}, \bibinfo {author} {\bibfnamefont {L.}~\bibnamefont
  {{Fidkowski}}}, \bibinfo {author} {\bibfnamefont {X.}~\bibnamefont {{Chen}}},
  \ and\ \bibinfo {author} {\bibfnamefont {A.}~\bibnamefont {{Vishwanath}}},\
  }\href@noop {} {\bibfield  {journal} {\bibinfo  {journal} {arXiv e-prints}\
  ,\ \bibinfo {eid} {arXiv:1406.3032}} (\bibinfo {year} {2014})},\ \Eprint
  {http://arxiv.org/abs/1406.3032} {arXiv:1406.3032 [cond-mat.str-el]}
  \BibitemShut {NoStop}%
\bibitem [{\citenamefont {{Barkeshli}}\ \emph {et~al.}(2016)\citenamefont
  {{Barkeshli}}, \citenamefont {{Bonderson}}, \citenamefont {{Jian}},
  \citenamefont {{Cheng}},\ and\ \citenamefont {{Walker}}}]{Maissam_2016}%
  \BibitemOpen
  \bibfield  {author} {\bibinfo {author} {\bibfnamefont {M.}~\bibnamefont
  {{Barkeshli}}}, \bibinfo {author} {\bibfnamefont {P.}~\bibnamefont
  {{Bonderson}}}, \bibinfo {author} {\bibfnamefont {C.-M.}\ \bibnamefont
  {{Jian}}}, \bibinfo {author} {\bibfnamefont {M.}~\bibnamefont {{Cheng}}}, \
  and\ \bibinfo {author} {\bibfnamefont {K.}~\bibnamefont {{Walker}}},\
  }\href@noop {} {\bibfield  {journal} {\bibinfo  {journal} {arXiv e-prints}\
  ,\ \bibinfo {eid} {arXiv:1612.07792}} (\bibinfo {year} {2016})},\ \Eprint
  {http://arxiv.org/abs/1612.07792} {arXiv:1612.07792 [cond-mat.str-el]}
  \BibitemShut {NoStop}%
\bibitem [{\citenamefont {Wang}\ and\ \citenamefont {Levin}(2017)}]{Wang_2017}%
  \BibitemOpen
  \bibfield  {author} {\bibinfo {author} {\bibfnamefont {C.}~\bibnamefont
  {Wang}}\ and\ \bibinfo {author} {\bibfnamefont {M.}~\bibnamefont {Levin}},\
  }\href {\doibase 10.1103/PhysRevLett.119.136801} {\bibfield  {journal}
  {\bibinfo  {journal} {Phys. Rev. Lett.}\ }\textbf {\bibinfo {volume} {119}},\
  \bibinfo {pages} {136801} (\bibinfo {year} {2017})}\BibitemShut {NoStop}%
\bibitem [{\citenamefont {{Tachikawa}}\ and\ \citenamefont
  {{Yonekura}}(2017)}]{Tachikawa_2017a}%
  \BibitemOpen
  \bibfield  {author} {\bibinfo {author} {\bibfnamefont {Y.}~\bibnamefont
  {{Tachikawa}}}\ and\ \bibinfo {author} {\bibfnamefont {K.}~\bibnamefont
  {{Yonekura}}},\ }\href {\doibase 10.1093/ptep/ptx010} {\bibfield  {journal}
  {\bibinfo  {journal} {Progress of Theoretical and Experimental Physics}\
  }\textbf {\bibinfo {volume} {2017}},\ \bibinfo {eid} {033B04} (\bibinfo
  {year} {2017})},\ \Eprint {http://arxiv.org/abs/1610.07010} {arXiv:1610.07010
  [hep-th]} \BibitemShut {NoStop}%
\bibitem [{\citenamefont {Tachikawa}\ and\ \citenamefont
  {Yonekura}(2017)}]{Tachikawa_2017}%
  \BibitemOpen
  \bibfield  {author} {\bibinfo {author} {\bibfnamefont {Y.}~\bibnamefont
  {Tachikawa}}\ and\ \bibinfo {author} {\bibfnamefont {K.}~\bibnamefont
  {Yonekura}},\ }\href {\doibase 10.1103/PhysRevLett.119.111603} {\bibfield
  {journal} {\bibinfo  {journal} {Phys. Rev. Lett.}\ }\textbf {\bibinfo
  {volume} {119}},\ \bibinfo {pages} {111603} (\bibinfo {year}
  {2017})}\BibitemShut {NoStop}%
\bibitem [{\citenamefont {Cheng}(2018)}]{Meng_2018}%
  \BibitemOpen
  \bibfield  {author} {\bibinfo {author} {\bibfnamefont {M.}~\bibnamefont
  {Cheng}},\ }\href {\doibase 10.1103/PhysRevLett.120.036801} {\bibfield
  {journal} {\bibinfo  {journal} {Phys. Rev. Lett.}\ }\textbf {\bibinfo
  {volume} {120}},\ \bibinfo {pages} {036801} (\bibinfo {year}
  {2018})}\BibitemShut {NoStop}%
\bibitem [{\citenamefont {Benalcazar}\ \emph
  {et~al.}(2017{\natexlab{a}})\citenamefont {Benalcazar}, \citenamefont
  {Bernevig},\ and\ \citenamefont {Hughes}}]{Benalcazar61}%
  \BibitemOpen
  \bibfield  {author} {\bibinfo {author} {\bibfnamefont {W.~A.}\ \bibnamefont
  {Benalcazar}}, \bibinfo {author} {\bibfnamefont {B.~A.}\ \bibnamefont
  {Bernevig}}, \ and\ \bibinfo {author} {\bibfnamefont {T.~L.}\ \bibnamefont
  {Hughes}},\ }\href {\doibase 10.1126/science.aah6442} {\bibfield  {journal}
  {\bibinfo  {journal} {Science}\ }\textbf {\bibinfo {volume} {357}},\ \bibinfo
  {pages} {61} (\bibinfo {year} {2017}{\natexlab{a}})}\BibitemShut {NoStop}%
\bibitem [{\citenamefont {Benalcazar}\ \emph
  {et~al.}(2017{\natexlab{b}})\citenamefont {Benalcazar}, \citenamefont
  {Bernevig},\ and\ \citenamefont {Hughes}}]{Wlad_2017}%
  \BibitemOpen
  \bibfield  {author} {\bibinfo {author} {\bibfnamefont {W.~A.}\ \bibnamefont
  {Benalcazar}}, \bibinfo {author} {\bibfnamefont {B.~A.}\ \bibnamefont
  {Bernevig}}, \ and\ \bibinfo {author} {\bibfnamefont {T.~L.}\ \bibnamefont
  {Hughes}},\ }\href {\doibase 10.1103/PhysRevB.96.245115} {\bibfield
  {journal} {\bibinfo  {journal} {Phys. Rev. B}\ }\textbf {\bibinfo {volume}
  {96}},\ \bibinfo {pages} {245115} (\bibinfo {year}
  {2017}{\natexlab{b}})}\BibitemShut {NoStop}%
\bibitem [{\citenamefont {Schindler}\ \emph
  {et~al.}(2018{\natexlab{a}})\citenamefont {Schindler}, \citenamefont {Cook},
  \citenamefont {Vergniory}, \citenamefont {Wang}, \citenamefont {Parkin},
  \citenamefont {Bernevig},\ and\ \citenamefont {Neupert}}]{Schindlereaat0346}%
  \BibitemOpen
  \bibfield  {author} {\bibinfo {author} {\bibfnamefont {F.}~\bibnamefont
  {Schindler}}, \bibinfo {author} {\bibfnamefont {A.~M.}\ \bibnamefont {Cook}},
  \bibinfo {author} {\bibfnamefont {M.~G.}\ \bibnamefont {Vergniory}}, \bibinfo
  {author} {\bibfnamefont {Z.}~\bibnamefont {Wang}}, \bibinfo {author}
  {\bibfnamefont {S.~S.~P.}\ \bibnamefont {Parkin}}, \bibinfo {author}
  {\bibfnamefont {B.~A.}\ \bibnamefont {Bernevig}}, \ and\ \bibinfo {author}
  {\bibfnamefont {T.}~\bibnamefont {Neupert}},\ }\href {\doibase
  10.1126/sciadv.aat0346} {\bibfield  {journal} {\bibinfo  {journal} {Science
  Advances}\ }\textbf {\bibinfo {volume} {4}} (\bibinfo {year}
  {2018}{\natexlab{a}}),\ 10.1126/sciadv.aat0346}\BibitemShut {NoStop}%
\bibitem [{\citenamefont {Song}\ \emph
  {et~al.}(2017{\natexlab{a}})\citenamefont {Song}, \citenamefont {Fang},\ and\
  \citenamefont {Fang}}]{PhysRevLett.119.246402}%
  \BibitemOpen
  \bibfield  {author} {\bibinfo {author} {\bibfnamefont {Z.}~\bibnamefont
  {Song}}, \bibinfo {author} {\bibfnamefont {Z.}~\bibnamefont {Fang}}, \ and\
  \bibinfo {author} {\bibfnamefont {C.}~\bibnamefont {Fang}},\ }\href {\doibase
  10.1103/PhysRevLett.119.246402} {\bibfield  {journal} {\bibinfo  {journal}
  {Phys. Rev. Lett.}\ }\textbf {\bibinfo {volume} {119}},\ \bibinfo {pages}
  {246402} (\bibinfo {year} {2017}{\natexlab{a}})}\BibitemShut {NoStop}%
\bibitem [{\citenamefont {Langbehn}\ \emph {et~al.}(2017)\citenamefont
  {Langbehn}, \citenamefont {Peng}, \citenamefont {Trifunovic}, \citenamefont
  {von Oppen},\ and\ \citenamefont {Brouwer}}]{Brouwer_et_al_2017}%
  \BibitemOpen
  \bibfield  {author} {\bibinfo {author} {\bibfnamefont {J.}~\bibnamefont
  {Langbehn}}, \bibinfo {author} {\bibfnamefont {Y.}~\bibnamefont {Peng}},
  \bibinfo {author} {\bibfnamefont {L.}~\bibnamefont {Trifunovic}}, \bibinfo
  {author} {\bibfnamefont {F.}~\bibnamefont {von Oppen}}, \ and\ \bibinfo
  {author} {\bibfnamefont {P.~W.}\ \bibnamefont {Brouwer}},\ }\href {\doibase
  10.1103/PhysRevLett.119.246401} {\bibfield  {journal} {\bibinfo  {journal}
  {Phys. Rev. Lett.}\ }\textbf {\bibinfo {volume} {119}},\ \bibinfo {pages}
  {246401} (\bibinfo {year} {2017})}\BibitemShut {NoStop}%
\bibitem [{\citenamefont {{S.A.~Parameswaran}}\ and\ \citenamefont
  {Wan}(2017)}]{parameswaran2017a}%
  \BibitemOpen
  \bibfield  {author} {\bibinfo {author} {\bibnamefont {{S.A.~Parameswaran}}}\
  and\ \bibinfo {author} {\bibfnamefont {Y.}~\bibnamefont {Wan}},\ }\href
  {https://physics.aps.org/articles/v10/132} {\bibfield  {journal} {\bibinfo
  {journal} {Physics}\ }\textbf {\bibinfo {volume} {10}},\ \bibinfo {pages} {1}
  (\bibinfo {year} {2017})}\BibitemShut {NoStop}%
\bibitem [{\citenamefont {Khalaf}(2018)}]{Khalaf_2018}%
  \BibitemOpen
  \bibfield  {author} {\bibinfo {author} {\bibfnamefont {E.}~\bibnamefont
  {Khalaf}},\ }\href {\doibase 10.1103/PhysRevB.97.205136} {\bibfield
  {journal} {\bibinfo  {journal} {Phys. Rev. B}\ }\textbf {\bibinfo {volume}
  {97}},\ \bibinfo {pages} {205136} (\bibinfo {year} {2018})}\BibitemShut
  {NoStop}%
\bibitem [{\citenamefont {Geier}\ \emph {et~al.}(2018)\citenamefont {Geier},
  \citenamefont {Trifunovic}, \citenamefont {Hoskam},\ and\ \citenamefont
  {Brouwer}}]{Brouwer_2018}%
  \BibitemOpen
  \bibfield  {author} {\bibinfo {author} {\bibfnamefont {M.}~\bibnamefont
  {Geier}}, \bibinfo {author} {\bibfnamefont {L.}~\bibnamefont {Trifunovic}},
  \bibinfo {author} {\bibfnamefont {M.}~\bibnamefont {Hoskam}}, \ and\ \bibinfo
  {author} {\bibfnamefont {P.~W.}\ \bibnamefont {Brouwer}},\ }\href {\doibase
  10.1103/PhysRevB.97.205135} {\bibfield  {journal} {\bibinfo  {journal} {Phys.
  Rev. B}\ }\textbf {\bibinfo {volume} {97}},\ \bibinfo {pages} {205135}
  (\bibinfo {year} {2018})}\BibitemShut {NoStop}%
\bibitem [{\citenamefont {Trifunovic}\ and\ \citenamefont
  {Brouwer}(2019)}]{Brouwer_2019}%
  \BibitemOpen
  \bibfield  {author} {\bibinfo {author} {\bibfnamefont {L.}~\bibnamefont
  {Trifunovic}}\ and\ \bibinfo {author} {\bibfnamefont {P.~W.}\ \bibnamefont
  {Brouwer}},\ }\href {\doibase 10.1103/PhysRevX.9.011012} {\bibfield
  {journal} {\bibinfo  {journal} {Phys. Rev. X}\ }\textbf {\bibinfo {volume}
  {9}},\ \bibinfo {pages} {011012} (\bibinfo {year} {2019})}\BibitemShut
  {NoStop}%
\bibitem [{\citenamefont {Schindler}\ \emph
  {et~al.}(2018{\natexlab{b}})\citenamefont {Schindler}, \citenamefont {Wang},
  \citenamefont {Vergniory}, \citenamefont {Cook}, \citenamefont {Murani},
  \citenamefont {Sengupta}, \citenamefont {Kasumov}, \citenamefont {Deblock},
  \citenamefont {Jeon}, \citenamefont {Drozdov}, \citenamefont {Bouchiat},
  \citenamefont {Gu{\'e}ron}, \citenamefont {Yazdani}, \citenamefont
  {Bernevig},\ and\ \citenamefont {Neupert}}]{schindler2018higher}%
  \BibitemOpen
  \bibfield  {author} {\bibinfo {author} {\bibfnamefont {F.}~\bibnamefont
  {Schindler}}, \bibinfo {author} {\bibfnamefont {Z.}~\bibnamefont {Wang}},
  \bibinfo {author} {\bibfnamefont {M.~G.}\ \bibnamefont {Vergniory}}, \bibinfo
  {author} {\bibfnamefont {A.~M.}\ \bibnamefont {Cook}}, \bibinfo {author}
  {\bibfnamefont {A.}~\bibnamefont {Murani}}, \bibinfo {author} {\bibfnamefont
  {S.}~\bibnamefont {Sengupta}}, \bibinfo {author} {\bibfnamefont {A.~Y.}\
  \bibnamefont {Kasumov}}, \bibinfo {author} {\bibfnamefont {R.}~\bibnamefont
  {Deblock}}, \bibinfo {author} {\bibfnamefont {S.}~\bibnamefont {Jeon}},
  \bibinfo {author} {\bibfnamefont {I.}~\bibnamefont {Drozdov}}, \bibinfo
  {author} {\bibfnamefont {H.}~\bibnamefont {Bouchiat}}, \bibinfo {author}
  {\bibfnamefont {S.}~\bibnamefont {Gu{\'e}ron}}, \bibinfo {author}
  {\bibfnamefont {A.}~\bibnamefont {Yazdani}}, \bibinfo {author} {\bibfnamefont
  {B.~A.}\ \bibnamefont {Bernevig}}, \ and\ \bibinfo {author} {\bibfnamefont
  {T.}~\bibnamefont {Neupert}},\ }\href {\doibase 10.1038/s41567-018-0224-7}
  {\bibfield  {journal} {\bibinfo  {journal} {{Nature Physics}}\ }\textbf
  {\bibinfo {volume} {14}},\ \bibinfo {pages} {918} (\bibinfo {year}
  {2018}{\natexlab{b}})}\BibitemShut {NoStop}%
\bibitem [{\citenamefont {Imhof}\ \emph {et~al.}(2018)\citenamefont {Imhof},
  \citenamefont {Berger}, \citenamefont {Bayer}, \citenamefont {Brehm},
  \citenamefont {Molenkamp}, \citenamefont {Kiessling}, \citenamefont
  {Schindler}, \citenamefont {Lee}, \citenamefont {Greiter}, \citenamefont
  {Neupert},\ and\ \citenamefont {Thomale}}]{imhof2018topolectrical}%
  \BibitemOpen
  \bibfield  {author} {\bibinfo {author} {\bibfnamefont {S.}~\bibnamefont
  {Imhof}}, \bibinfo {author} {\bibfnamefont {C.}~\bibnamefont {Berger}},
  \bibinfo {author} {\bibfnamefont {F.}~\bibnamefont {Bayer}}, \bibinfo
  {author} {\bibfnamefont {J.}~\bibnamefont {Brehm}}, \bibinfo {author}
  {\bibfnamefont {L.~W.}\ \bibnamefont {Molenkamp}}, \bibinfo {author}
  {\bibfnamefont {T.}~\bibnamefont {Kiessling}}, \bibinfo {author}
  {\bibfnamefont {F.}~\bibnamefont {Schindler}}, \bibinfo {author}
  {\bibfnamefont {C.~H.}\ \bibnamefont {Lee}}, \bibinfo {author} {\bibfnamefont
  {M.}~\bibnamefont {Greiter}}, \bibinfo {author} {\bibfnamefont
  {T.}~\bibnamefont {Neupert}}, \ and\ \bibinfo {author} {\bibfnamefont
  {R.}~\bibnamefont {Thomale}},\ }\href {\doibase 10.1038/s41567-018-0246-1}
  {\bibfield  {journal} {\bibinfo  {journal} {Nature Physics}\ }\textbf
  {\bibinfo {volume} {14}},\ \bibinfo {pages} {925} (\bibinfo {year}
  {2018})}\BibitemShut {NoStop}%
\bibitem [{\citenamefont {You}\ \emph {et~al.}(2018)\citenamefont {You},
  \citenamefont {Devakul}, \citenamefont {Burnell},\ and\ \citenamefont
  {Neupert}}]{Yizhi_2018}%
  \BibitemOpen
  \bibfield  {author} {\bibinfo {author} {\bibfnamefont {Y.}~\bibnamefont
  {You}}, \bibinfo {author} {\bibfnamefont {T.}~\bibnamefont {Devakul}},
  \bibinfo {author} {\bibfnamefont {F.~J.}\ \bibnamefont {Burnell}}, \ and\
  \bibinfo {author} {\bibfnamefont {T.}~\bibnamefont {Neupert}},\ }\href
  {\doibase 10.1103/PhysRevB.98.235102} {\bibfield  {journal} {\bibinfo
  {journal} {Phys. Rev. B}\ }\textbf {\bibinfo {volume} {98}},\ \bibinfo
  {pages} {235102} (\bibinfo {year} {2018})}\BibitemShut {NoStop}%
\bibitem [{\citenamefont {{Rasmussen}}\ and\ \citenamefont
  {{Lu}}(2018)}]{Rasmussen_Lu_2018}%
  \BibitemOpen
  \bibfield  {author} {\bibinfo {author} {\bibfnamefont {A.}~\bibnamefont
  {{Rasmussen}}}\ and\ \bibinfo {author} {\bibfnamefont {Y.-M.}\ \bibnamefont
  {{Lu}}},\ }\href@noop {} {\bibfield  {journal} {\bibinfo  {journal} {arXiv
  e-prints}\ ,\ \bibinfo {eid} {arXiv:1809.07325}} (\bibinfo {year} {2018})},\
  \Eprint {http://arxiv.org/abs/1809.07325} {arXiv:1809.07325
  [cond-mat.str-el]} \BibitemShut {NoStop}%
\bibitem [{\citenamefont {{Ghorashi}}\ \emph {et~al.}(2019)\citenamefont
  {{Ghorashi}}, \citenamefont {{Hu}}, \citenamefont {{Hughes}},\ and\
  \citenamefont {{Rossi}}}]{2019arXiv190107579G}%
  \BibitemOpen
  \bibfield  {author} {\bibinfo {author} {\bibfnamefont {S.~A.~A.}\
  \bibnamefont {{Ghorashi}}}, \bibinfo {author} {\bibfnamefont
  {X.}~\bibnamefont {{Hu}}}, \bibinfo {author} {\bibfnamefont {T.~L.}\
  \bibnamefont {{Hughes}}}, \ and\ \bibinfo {author} {\bibfnamefont
  {E.}~\bibnamefont {{Rossi}}},\ }\href@noop {} {\bibfield  {journal} {\bibinfo
   {journal} {arXiv e-prints}\ ,\ \bibinfo {eid} {arXiv:1901.07579}} (\bibinfo
  {year} {2019})},\ \Eprint {http://arxiv.org/abs/1901.07579} {arXiv:1901.07579
  [cond-mat.mes-hall]} \BibitemShut {NoStop}%
\bibitem [{\citenamefont {Song}\ \emph
  {et~al.}(2017{\natexlab{b}})\citenamefont {Song}, \citenamefont {Huang},
  \citenamefont {Fu},\ and\ \citenamefont {Hermele}}]{PhysRevX.7.011020}%
  \BibitemOpen
  \bibfield  {author} {\bibinfo {author} {\bibfnamefont {H.}~\bibnamefont
  {Song}}, \bibinfo {author} {\bibfnamefont {S.-J.}\ \bibnamefont {Huang}},
  \bibinfo {author} {\bibfnamefont {L.}~\bibnamefont {Fu}}, \ and\ \bibinfo
  {author} {\bibfnamefont {M.}~\bibnamefont {Hermele}},\ }\href {\doibase
  10.1103/PhysRevX.7.011020} {\bibfield  {journal} {\bibinfo  {journal} {Phys.
  Rev. X}\ }\textbf {\bibinfo {volume} {7}},\ \bibinfo {pages} {011020}
  (\bibinfo {year} {2017}{\natexlab{b}})}\BibitemShut {NoStop}%
\bibitem [{\citenamefont {Isobe}\ and\ \citenamefont
  {Fu}(2015)}]{PhysRevB.92.081304}%
  \BibitemOpen
  \bibfield  {author} {\bibinfo {author} {\bibfnamefont {H.}~\bibnamefont
  {Isobe}}\ and\ \bibinfo {author} {\bibfnamefont {L.}~\bibnamefont {Fu}},\
  }\href {\doibase 10.1103/PhysRevB.92.081304} {\bibfield  {journal} {\bibinfo
  {journal} {Phys. Rev. B}\ }\textbf {\bibinfo {volume} {92}},\ \bibinfo
  {pages} {081304} (\bibinfo {year} {2015})}\BibitemShut {NoStop}%
\bibitem [{\citenamefont {Jiang}\ and\ \citenamefont
  {Ran}(2017)}]{PhysRevB.95.125107}%
  \BibitemOpen
  \bibfield  {author} {\bibinfo {author} {\bibfnamefont {S.}~\bibnamefont
  {Jiang}}\ and\ \bibinfo {author} {\bibfnamefont {Y.}~\bibnamefont {Ran}},\
  }\href {\doibase 10.1103/PhysRevB.95.125107} {\bibfield  {journal} {\bibinfo
  {journal} {Phys. Rev. B}\ }\textbf {\bibinfo {volume} {95}},\ \bibinfo
  {pages} {125107} (\bibinfo {year} {2017})}\BibitemShut {NoStop}%
\bibitem [{\citenamefont {Huang}\ \emph {et~al.}(2017)\citenamefont {Huang},
  \citenamefont {Song}, \citenamefont {Huang},\ and\ \citenamefont
  {Hermele}}]{PhysRevB.96.205106}%
  \BibitemOpen
  \bibfield  {author} {\bibinfo {author} {\bibfnamefont {S.-J.}\ \bibnamefont
  {Huang}}, \bibinfo {author} {\bibfnamefont {H.}~\bibnamefont {Song}},
  \bibinfo {author} {\bibfnamefont {Y.-P.}\ \bibnamefont {Huang}}, \ and\
  \bibinfo {author} {\bibfnamefont {M.}~\bibnamefont {Hermele}},\ }\href
  {\doibase 10.1103/PhysRevB.96.205106} {\bibfield  {journal} {\bibinfo
  {journal} {Phys. Rev. B}\ }\textbf {\bibinfo {volume} {96}},\ \bibinfo
  {pages} {205106} (\bibinfo {year} {2017})}\BibitemShut {NoStop}%
\bibitem [{\citenamefont {Thorngren}\ and\ \citenamefont
  {Else}(2018)}]{PhysRevX.8.011040}%
  \BibitemOpen
  \bibfield  {author} {\bibinfo {author} {\bibfnamefont {R.}~\bibnamefont
  {Thorngren}}\ and\ \bibinfo {author} {\bibfnamefont {D.~V.}\ \bibnamefont
  {Else}},\ }\href {\doibase 10.1103/PhysRevX.8.011040} {\bibfield  {journal}
  {\bibinfo  {journal} {Phys. Rev. X}\ }\textbf {\bibinfo {volume} {8}},\
  \bibinfo {pages} {011040} (\bibinfo {year} {2018})}\BibitemShut {NoStop}%
\bibitem [{\citenamefont {{Kobayashi}}\ and\ \citenamefont
  {{Shiozaki}}(2019)}]{2019arXiv190106195K}%
  \BibitemOpen
  \bibfield  {author} {\bibinfo {author} {\bibfnamefont {R.}~\bibnamefont
  {{Kobayashi}}}\ and\ \bibinfo {author} {\bibfnamefont {K.}~\bibnamefont
  {{Shiozaki}}},\ }\href@noop {} {\bibfield  {journal} {\bibinfo  {journal}
  {arXiv e-prints}\ ,\ \bibinfo {eid} {arXiv:1901.06195}} (\bibinfo {year}
  {2019})},\ \Eprint {http://arxiv.org/abs/1901.06195} {arXiv:1901.06195
  [cond-mat.str-el]} \BibitemShut {NoStop}%
\bibitem [{\citenamefont {Francesco}\ \emph {et~al.}(2012)\citenamefont
  {Francesco}, \citenamefont {Mathieu},\ and\ \citenamefont
  {S{\'e}n{\'e}chal}}]{francesco2012conformal}%
  \BibitemOpen
  \bibfield  {author} {\bibinfo {author} {\bibfnamefont {P.}~\bibnamefont
  {Francesco}}, \bibinfo {author} {\bibfnamefont {P.}~\bibnamefont {Mathieu}},
  \ and\ \bibinfo {author} {\bibfnamefont {D.}~\bibnamefont
  {S{\'e}n{\'e}chal}},\ }\href@noop {} {\emph {\bibinfo {title} {Conformal
  field theory}}}\ (\bibinfo  {publisher} {Springer Science \& Business
  Media},\ \bibinfo {year} {2012})\BibitemShut {NoStop}%
\bibitem [{\citenamefont {Fradkin}\ \emph {et~al.}(1998)\citenamefont
  {Fradkin}, \citenamefont {Nayak}, \citenamefont {Tsvelik},\ and\
  \citenamefont {Wilczek}}]{Fradkin_98}%
  \BibitemOpen
  \bibfield  {author} {\bibinfo {author} {\bibfnamefont {E.}~\bibnamefont
  {Fradkin}}, \bibinfo {author} {\bibfnamefont {C.}~\bibnamefont {Nayak}},
  \bibinfo {author} {\bibfnamefont {A.}~\bibnamefont {Tsvelik}}, \ and\
  \bibinfo {author} {\bibfnamefont {F.}~\bibnamefont {Wilczek}},\ }\href
  {\doibase https://doi.org/10.1016/S0550-3213(98)00111-4} {\bibfield
  {journal} {\bibinfo  {journal} {Nuclear Physics B}\ }\textbf {\bibinfo
  {volume} {516}},\ \bibinfo {pages} {704 } (\bibinfo {year}
  {1998})}\BibitemShut {NoStop}%
\bibitem [{\citenamefont {Fendley}\ \emph {et~al.}(2007)\citenamefont
  {Fendley}, \citenamefont {Fisher},\ and\ \citenamefont
  {Nayak}}]{Fendley_2007}%
  \BibitemOpen
  \bibfield  {author} {\bibinfo {author} {\bibfnamefont {P.}~\bibnamefont
  {Fendley}}, \bibinfo {author} {\bibfnamefont {M.~P.~A.}\ \bibnamefont
  {Fisher}}, \ and\ \bibinfo {author} {\bibfnamefont {C.}~\bibnamefont
  {Nayak}},\ }\href {\doibase 10.1103/PhysRevB.75.045317} {\bibfield  {journal}
  {\bibinfo  {journal} {Phys. Rev. B}\ }\textbf {\bibinfo {volume} {75}},\
  \bibinfo {pages} {045317} (\bibinfo {year} {2007})}\BibitemShut {NoStop}%
\bibitem [{\citenamefont {Bishara}\ and\ \citenamefont
  {Nayak}(2008)}]{Bishara_2008}%
  \BibitemOpen
  \bibfield  {author} {\bibinfo {author} {\bibfnamefont {W.}~\bibnamefont
  {Bishara}}\ and\ \bibinfo {author} {\bibfnamefont {C.}~\bibnamefont
  {Nayak}},\ }\href {\doibase 10.1103/PhysRevB.77.165302} {\bibfield  {journal}
  {\bibinfo  {journal} {Phys. Rev. B}\ }\textbf {\bibinfo {volume} {77}},\
  \bibinfo {pages} {165302} (\bibinfo {year} {2008})}\BibitemShut {NoStop}%
\bibitem [{Sup()}]{SupMat}%
  \BibitemOpen
  \href@noop {} {}\bibinfo {note} {{See supplementary material for (i)
  discussion of gauge symmetry in the $\mathcal{T}$-Pfaffian and (ii) details
  of anyon condensation in HOTI and HOTSC.}}\BibitemShut {Stop}%
\bibitem [{\citenamefont {Bruillard}\ \emph {et~al.}(2017)\citenamefont
  {Bruillard}, \citenamefont {Galindo}, \citenamefont {Hagge}, \citenamefont
  {Ng}, \citenamefont {Plavnik}, \citenamefont {Rowell},\ and\ \citenamefont
  {Wang}}]{Rubenstein_2016}%
  \BibitemOpen
  \bibfield  {author} {\bibinfo {author} {\bibfnamefont {P.}~\bibnamefont
  {Bruillard}}, \bibinfo {author} {\bibfnamefont {C.}~\bibnamefont {Galindo}},
  \bibinfo {author} {\bibfnamefont {T.}~\bibnamefont {Hagge}}, \bibinfo
  {author} {\bibfnamefont {S.-H.}\ \bibnamefont {Ng}}, \bibinfo {author}
  {\bibfnamefont {J.~Y.}\ \bibnamefont {Plavnik}}, \bibinfo {author}
  {\bibfnamefont {E.~C.}\ \bibnamefont {Rowell}}, \ and\ \bibinfo {author}
  {\bibfnamefont {Z.}~\bibnamefont {Wang}},\ }\href {\doibase
  10.1063/1.4982048} {\bibfield  {journal} {\bibinfo  {journal} {Journal of
  Mathematical Physics}\ }\textbf {\bibinfo {volume} {58}},\ \bibinfo {pages}
  {041704} (\bibinfo {year} {2017})},\ \Eprint
  {http://arxiv.org/abs/https://doi.org/10.1063/1.4982048}
  {https://doi.org/10.1063/1.4982048} \BibitemShut {NoStop}%
\bibitem [{\citenamefont {Bhardwaj}\ \emph {et~al.}(2017)\citenamefont
  {Bhardwaj}, \citenamefont {Gaiotto},\ and\ \citenamefont
  {Kapustin}}]{Bhardwaj2017}%
  \BibitemOpen
  \bibfield  {author} {\bibinfo {author} {\bibfnamefont {L.}~\bibnamefont
  {Bhardwaj}}, \bibinfo {author} {\bibfnamefont {D.}~\bibnamefont {Gaiotto}}, \
  and\ \bibinfo {author} {\bibfnamefont {A.}~\bibnamefont {Kapustin}},\ }\href
  {\doibase 10.1007/JHEP04(2017)096} {\bibfield  {journal} {\bibinfo  {journal}
  {Journal of High Energy Physics}\ }\textbf {\bibinfo {volume} {2017}},\
  \bibinfo {pages} {96} (\bibinfo {year} {2017})}\BibitemShut {NoStop}%
\bibitem [{\citenamefont {Aasen}\ \emph {et~al.}(2017)\citenamefont {Aasen},
  \citenamefont {Lake},\ and\ \citenamefont {Walker}}]{aasen2017}%
  \BibitemOpen
  \bibfield  {author} {\bibinfo {author} {\bibfnamefont {D.}~\bibnamefont
  {Aasen}}, \bibinfo {author} {\bibfnamefont {E.}~\bibnamefont {Lake}}, \ and\
  \bibinfo {author} {\bibfnamefont {K.}~\bibnamefont {Walker}},\ }\href@noop {}
  {\bibfield  {journal} {\bibinfo  {journal} {arXiv preprint arXiv:1709.01941}\
  } (\bibinfo {year} {2017})}\BibitemShut {NoStop}%
\bibitem [{Note1()}]{Note1}%
  \BibitemOpen
  \bibinfo {note} {$\protect \mathcal {A}_{\protect \mathcal {T}}$ is the
  `$\protect \mathcal {T}$-Pfaffian$_+$', the STO of the free-fermion TI. A
  distinct `$\protect \mathcal {T}$-Pfaffian$_-$' with sign-reversed
  topological spins and $\protect \mathcal {T}^2$ actions (where defined) for
  $1_{2,6},\psi _{2,6}$ and $\sigma _j$ yields an STO for an intrinsically
  interacting HOTI.}\BibitemShut {Stop}%
\bibitem [{\citenamefont {Haldane}(1995)}]{Haldane_1995}%
  \BibitemOpen
  \bibfield  {author} {\bibinfo {author} {\bibfnamefont {F.~D.~M.}\
  \bibnamefont {Haldane}},\ }\href {\doibase 10.1103/PhysRevLett.74.2090}
  {\bibfield  {journal} {\bibinfo  {journal} {Phys. Rev. Lett.}\ }\textbf
  {\bibinfo {volume} {74}},\ \bibinfo {pages} {2090} (\bibinfo {year}
  {1995})}\BibitemShut {NoStop}%
\bibitem [{\citenamefont {Levin}(2013)}]{Levin_2013}%
  \BibitemOpen
  \bibfield  {author} {\bibinfo {author} {\bibfnamefont {M.}~\bibnamefont
  {Levin}},\ }\href {\doibase 10.1103/PhysRevX.3.021009} {\bibfield  {journal}
  {\bibinfo  {journal} {Phys. Rev. X}\ }\textbf {\bibinfo {volume} {3}},\
  \bibinfo {pages} {021009} (\bibinfo {year} {2013})}\BibitemShut {NoStop}%
\bibitem [{\citenamefont {Wang}\ and\ \citenamefont {Wen}(2015)}]{Juven_2015}%
  \BibitemOpen
  \bibfield  {author} {\bibinfo {author} {\bibfnamefont {J.~C.}\ \bibnamefont
  {Wang}}\ and\ \bibinfo {author} {\bibfnamefont {X.-G.}\ \bibnamefont {Wen}},\
  }\href {\doibase 10.1103/PhysRevB.91.125124} {\bibfield  {journal} {\bibinfo
  {journal} {Phys. Rev. B}\ }\textbf {\bibinfo {volume} {91}},\ \bibinfo
  {pages} {125124} (\bibinfo {year} {2015})}\BibitemShut {NoStop}%
\bibitem [{\citenamefont {Tiwari}\ \emph {et~al.}(2019)\citenamefont {Tiwari},
  \citenamefont {Li}, \citenamefont {Neupert}, \citenamefont {{B.A.
  Bernevig}},\ and\ \citenamefont {{S.A.~Parameswaran}}}]{TLNBP-unpub}%
  \BibitemOpen
  \bibfield  {author} {\bibinfo {author} {\bibfnamefont {A.}~\bibnamefont
  {Tiwari}}, \bibinfo {author} {\bibfnamefont {M.}~\bibnamefont {Li}}, \bibinfo
  {author} {\bibfnamefont {T.}~\bibnamefont {Neupert}}, \bibinfo {author}
  {\bibnamefont {{B.A. Bernevig}}}, \ and\ \bibinfo {author} {\bibnamefont
  {{S.A.~Parameswaran}}},\ }\href@noop {} {\enquote {\bibinfo {title} {Gapped
  surfaces for helical higher-order topological phases},}\ } (\bibinfo {year}
  {2019}),\ \bibinfo {note} {in preparation}\BibitemShut {NoStop}%
\bibitem [{Note2()}]{Note2}%
  \BibitemOpen
  \bibinfo {note} {We thank Meng Cheng for discussions on this
  point.}\BibitemShut {Stop}%
\bibitem [{\citenamefont {Bais}\ and\ \citenamefont
  {Slingerland}(2009)}]{PhysRevB.79.045316}%
  \BibitemOpen
  \bibfield  {author} {\bibinfo {author} {\bibfnamefont {F.~A.}\ \bibnamefont
  {Bais}}\ and\ \bibinfo {author} {\bibfnamefont {J.~K.}\ \bibnamefont
  {Slingerland}},\ }\href {\doibase 10.1103/PhysRevB.79.045316} {\bibfield
  {journal} {\bibinfo  {journal} {Phys. Rev. B}\ }\textbf {\bibinfo {volume}
  {79}},\ \bibinfo {pages} {045316} (\bibinfo {year} {2009})}\BibitemShut
  {NoStop}%
\bibitem [{\citenamefont {Kong}(2014)}]{KONG2014436}%
  \BibitemOpen
  \bibfield  {author} {\bibinfo {author} {\bibfnamefont {L.}~\bibnamefont
  {Kong}},\ }\href {\doibase https://doi.org/10.1016/j.nuclphysb.2014.07.003}
  {\bibfield  {journal} {\bibinfo  {journal} {Nuclear Physics B}\ }\textbf
  {\bibinfo {volume} {886}},\ \bibinfo {pages} {436 } (\bibinfo {year}
  {2014})}\BibitemShut {NoStop}%
\bibitem [{\citenamefont {Neupert}\ \emph {et~al.}(2016)\citenamefont
  {Neupert}, \citenamefont {He}, \citenamefont {von Keyserlingk}, \citenamefont
  {Sierra},\ and\ \citenamefont {Bernevig}}]{Titus_Boson_2016}%
  \BibitemOpen
  \bibfield  {author} {\bibinfo {author} {\bibfnamefont {T.}~\bibnamefont
  {Neupert}}, \bibinfo {author} {\bibfnamefont {H.}~\bibnamefont {He}},
  \bibinfo {author} {\bibfnamefont {C.}~\bibnamefont {von Keyserlingk}},
  \bibinfo {author} {\bibfnamefont {G.}~\bibnamefont {Sierra}}, \ and\ \bibinfo
  {author} {\bibfnamefont {B.~A.}\ \bibnamefont {Bernevig}},\ }\href {\doibase
  10.1103/PhysRevB.93.115103} {\bibfield  {journal} {\bibinfo  {journal} {Phys.
  Rev. B}\ }\textbf {\bibinfo {volume} {93}},\ \bibinfo {pages} {115103}
  (\bibinfo {year} {2016})}\BibitemShut {NoStop}%
\bibitem [{\citenamefont {{Kapustin}}\ and\ \citenamefont
  {{Saulina}}(2010)}]{Kapustin_2010}%
  \BibitemOpen
  \bibfield  {author} {\bibinfo {author} {\bibfnamefont {A.}~\bibnamefont
  {{Kapustin}}}\ and\ \bibinfo {author} {\bibfnamefont {N.}~\bibnamefont
  {{Saulina}}},\ }\href@noop {} {\bibfield  {journal} {\bibinfo  {journal}
  {arXiv e-prints}\ ,\ \bibinfo {eid} {arXiv:1012.0911}} (\bibinfo {year}
  {2010})},\ \Eprint {http://arxiv.org/abs/1012.0911} {arXiv:1012.0911
  [hep-th]} \BibitemShut {NoStop}%
\bibitem [{\citenamefont {Bakalov}\ and\ \citenamefont
  {Kirillov}(2001)}]{bakalov2001lectures}%
  \BibitemOpen
  \bibfield  {author} {\bibinfo {author} {\bibfnamefont {B.}~\bibnamefont
  {Bakalov}}\ and\ \bibinfo {author} {\bibfnamefont {A.~A.}\ \bibnamefont
  {Kirillov}},\ }\href@noop {} {\emph {\bibinfo {title} {Lectures on tensor
  categories and modular functors}}},\ Vol.~\bibinfo {volume} {21}\ (\bibinfo
  {publisher} {American Mathematical Soc.},\ \bibinfo {year}
  {2001})\BibitemShut {NoStop}%
\bibitem [{\citenamefont {Moore}\ and\ \citenamefont
  {Seiberg}(1990)}]{moore1990lectures}%
  \BibitemOpen
  \bibfield  {author} {\bibinfo {author} {\bibfnamefont {G.}~\bibnamefont
  {Moore}}\ and\ \bibinfo {author} {\bibfnamefont {N.}~\bibnamefont
  {Seiberg}},\ }in\ \href@noop {} {\emph {\bibinfo {booktitle} {Physics,
  geometry and topology}}}\ (\bibinfo  {publisher} {Springer},\ \bibinfo {year}
  {1990})\ pp.\ \bibinfo {pages} {263--361}\BibitemShut {NoStop}%
\bibitem [{\citenamefont {{Ginsparg}}(1988)}]{Ginsparg_CFT}%
  \BibitemOpen
  \bibfield  {author} {\bibinfo {author} {\bibfnamefont {P.}~\bibnamefont
  {{Ginsparg}}},\ }\href@noop {} {\bibfield  {journal} {\bibinfo  {journal}
  {arXiv e-prints}\ ,\ \bibinfo {eid} {hep-th/9108028}} (\bibinfo {year}
  {1988})},\ \Eprint {http://arxiv.org/abs/hep-th/9108028}
  {arXiv:hep-th/9108028 [hep-th]} \BibitemShut {NoStop}%
\bibitem [{Note3()}]{Note3}%
  \BibitemOpen
  \bibinfo {note} {Sometimes also referred to as $\protect \mathsf
  {PSU}(2)_6$.}\BibitemShut {Stop}%
\bibitem [{\citenamefont {Verlinde}(1988)}]{Verlinde_1988}%
  \BibitemOpen
  \bibfield  {author} {\bibinfo {author} {\bibfnamefont {E.}~\bibnamefont
  {Verlinde}},\ }\href {\doibase https://doi.org/10.1016/0550-3213(88)90603-7}
  {\bibfield  {journal} {\bibinfo  {journal} {Nuclear Physics B}\ }\textbf
  {\bibinfo {volume} {300}},\ \bibinfo {pages} {360 } (\bibinfo {year}
  {1988})}\BibitemShut {NoStop}%
\bibitem [{\citenamefont
  {{Kapustin}}(2014{\natexlab{a}})}]{2014arXiv1403.1467K}%
  \BibitemOpen
  \bibfield  {author} {\bibinfo {author} {\bibfnamefont {A.}~\bibnamefont
  {{Kapustin}}},\ }\href@noop {} {\bibfield  {journal} {\bibinfo  {journal}
  {arXiv e-prints}\ ,\ \bibinfo {eid} {arXiv:1403.1467}} (\bibinfo {year}
  {2014}{\natexlab{a}})},\ \Eprint {http://arxiv.org/abs/1403.1467}
  {arXiv:1403.1467 [cond-mat.str-el]} \BibitemShut {NoStop}%
\bibitem [{\citenamefont
  {{Kapustin}}(2014{\natexlab{b}})}]{2014arXiv1404.6659K}%
  \BibitemOpen
  \bibfield  {author} {\bibinfo {author} {\bibfnamefont {A.}~\bibnamefont
  {{Kapustin}}},\ }\href@noop {} {\bibfield  {journal} {\bibinfo  {journal}
  {arXiv e-prints}\ ,\ \bibinfo {eid} {arXiv:1404.6659}} (\bibinfo {year}
  {2014}{\natexlab{b}})},\ \Eprint {http://arxiv.org/abs/1404.6659}
  {arXiv:1404.6659 [cond-mat.str-el]} \BibitemShut {NoStop}%
\bibitem [{Note4()}]{Note4}%
  \BibitemOpen
  \bibinfo {note} {An invertible topological order is one with no
  fractionalized bulk excitations and possibly a non-zero chiral central
  charge. Examples include the $\protect \mathsf {E}_8$ state of bosons, the
  integer quantum Hall effect, and the $\protect \mathsf {p} + i\protect
  \mathsf {p}$ superconductor.}\BibitemShut {Stop}%
\end{thebibliography}%
\begin{widetext}
\appendix
\section{Gauging, Condensation, and the $\mathcal T$-Pfaffian}
We comment briefly on the origin and role of gauge symmetry in the edge theory~\footnote{We thank Meng Cheng for discussions on this point.}. On a physical level it serves to identify the appropriate combination of neutral Majorana and charged $\ms{U}(1)$ boson fields that corresponds to the physical electron operator in the edge theory. This is also consistent with the identification of $\psi_4$ in the bulk as the physical electron --- recall that this procedure restricted the types of allowed bulk anyons. Analogously, it constrains the operators in the edge theory. 

More generally, gauging/orbifolding by a finite abelian group and edge condensation can be seen as `dual' processes. Starting from a topological order $\mathcal A$ with a global $\ms{G}$ symmetry where $\ms{G}$ is a finite abelian group, one may gauge $\ms{G}$ to obtain a new larger topological order $\mathcal A_{/\ms{G}}$. This increases the quantum dimension of the theory i.e $d_{\mathcal A_{/\ms{G}}}=|\ms{G}|d_{\mathcal A}$. Some examples of the corresponding edge phenomena of relevance to the present work are (i) gauging $\phi^{\ms{a}} \to \phi^{\ms{a}} +\pi/2$ in the $\ms{U}(1)_2$ CFT (which may be also viewed as a redefinition of the fundamental $\ms{U}(1)$ charge or the compactification radius of the edge CFT) furnishes the $\ms{U}(1)_8$ CFT and (ii) gauging $\psi^{\ms{a}}\to -\psi^{\ms{a}}$ in free Majorana CFT furnishes the Ising CFT. Conversely, one may start from a larger theory and condense a set of bosons $\mathcal B$ to obtain a theory with a smaller quantum dimension. Interpreting $\mathcal B$ as an abelian group with fusion providing the group multiplication structure, the smaller condensed theory has a global $\mathcal B$ symmetry. Let us revisit the above two examples in this light: (i) Starting from $\ms{U}(1)_8$ CFT, one may obtain $\ms{U}(1)_2$ by condensing the $j=4$ operator and (ii) starting from the Ising CFT, one may obtain a Majorana CFT by (fermion) condensing $\psi$ (i.e., by binding to a physical electron and then condensing). 

\begin{figure}[h!]
\centering
\includegraphics[width=0.4\columnwidth]{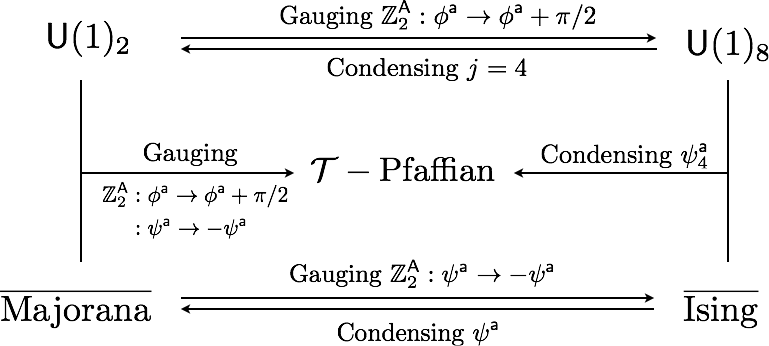}
\caption{The edge theory of the $\mathcal T$-Pfaffian topological order may be obtained by condensing $\psi^{\ms{a}}_4\in \ms{U}(1)_8\times \overline{\text{Ising}}$ or conversely by gauging a diagonal $\mathbb Z^{\ms{A}}_{2}$ symmetry in $U(1)_{2}\times \overline{\text{Majorana}}$ CFT.}
\label{fig:Z_2gauge}
\end{figure}

This gives us two ways to think of the $\mathcal T$-Pfaffian topological order. The condensation route considers the product of two modular tensor categories $\overline{\text{Ising}}\times \ms{U}(1)_8$  and  yields a non-modular theory by condensing the composite object $\psi_4\in \ms{U}(1)_8\times \overline{\text{Ising}}$, which essentially is built by binding the $4e/8$ charge anyon $\ms{U}(1)_8$  to the neutral fermion in $\overline{\text{Ising}}$ (again, we remind the reader that this condensation is implemented by identifying this object with the physical electron and then building a bilinear which is a boson and can thus be condensed). This confines a subset of the anyons in the product theory, leaving the (non-modular) $\mathcal T$-Pfaffian. Note that the object that is condensed is a composite of anyons in {\it both} $\ms{U}(1)_8$ and $\overline{\text{Ising}}$, so the resulting theory is not a simple product of topological orders accessible from either individually.  Another route starts the prooduct intrinsic invertible topological order $\overline{\text{Majorana}}$ (equivalent to the $\ms{p}-i\ms{p}$ superconductor) and the simpler Abelian theory $U(1)_{2}$. We then `gauge' the $\mathbb{Z}_2$ symmetry given by the product of fermion parity and the boson number conservation modulo 2. Gauging these symmetries independently would yield $\overline{\text{Ising}}\times \ms{U}(1)_8$, but gauging only their product yields the $\mathcal T$-Pfaffian. In other words, the middle line of Fig.~\ref{fig:Z_2gauge} indicates the equivalence of two approaches: namely (i) gauging both the $\mathbb{Z}_2$'s (fermion parity and boson number mod 2) and then `Higgsing'  (condensing) a $\mathbb{Z}_2$ subgroup, versus (ii) gauging only the diagonal $\mathbb{Z}_2$ subgroup  at the outset. 

Of these, the latter construction is a more convenient way to describe the edge theory, which is thus described by Eq.(1) of the main text augmented with the following gauge symmetry:
\begin{align}
\mathbb Z^{\ms{a}}_{2}: \psi^{\ms{a}} \mapsto -\psi^{\ms{a}} , \,\,\,\,\, \phi^{\ms{a}}\mapsto \phi^{\ms{a}} + \frac{\pi}{2} \,\,\,\,\text{with} \,\,\,\, \ms{a} = \ms{A}, \bar{\ms{A}}.
\end{align}
We now explain how the symmetry is enforced in our gapping perturbations on the doubled theory $\mathcal{L} = \mathcal{L}_{-}^{\mathsf A} + \mathcal{L}_{+}^{\bar{\mathsf A}}$ described in the main text. Since the compact boson $\phi$ is defined via $e^{i\phi}\sim \psi^{\ms{a}}+i \psi^{\bar{\ms{a}}}$, the action of the $\mathbb Z_{2}^{\ms{A},\bar{\ms{A}}}$ on the bosonic field $\phi$ and the chiral fields $\phi^{\ms{A}}, \phi^{\bar{\ms{A}}}$ can be written as
\begin{align}
\mathbb Z_{2}^{\ms{A}}:& &\phi \mapsto -\phi + \pi, & & \phi^{\ms{A}}\mapsto \phi^{\ms{A}} + \frac{\pi}{2}; \nonumber \\
\mathbb Z_{2}^{\bar{\ms{A}}}:& &\phi \mapsto -\phi, & & \phi^{\bar{\ms{A}}} \mapsto \phi^{\bar{\ms{A}}}  + \frac{\pi}{2}.
\end{align}
By adding the gapping term corresponding to the vector $\ell_{1}^{\ms{T}}=(0,4,4,4)$, the groundstate acquires a definite value for the field $\phi^{\ms{A}}+\phi^{\bar{\ms{A}}}+ \varphi$. Note that while naively the cosine term corresponding to $\ell_1$ apparently has four independent minima for any $\phi^{\ms{A}}+\phi^{\bar{\ms{A}}}+ \varphi = k \pi/2$ with $k = 0,1,2,3$, we can relate these by the gauge transformations, so there is only one unique minimum; we can use the gauge freedom to, e.g.,  fix  $\langle \phi^{\ms{A}}+\phi^{\bar{\ms{A}}}+ \varphi \rangle =0$. Thereafter only a subset of the $\mathbb Z_{2}^{\ms{A}}\times \mathbb Z_{2}^{\bar{\ms{A}}}$ transformations that leave this field combination invariant survives:  the corresponding group  is denoted $\widetilde{\mathbb Z}_{2}$ and acts as
\begin{align} 
\widetilde{\mathbb Z}_{2}:&\phi\to \phi+\pi, & & \phi^{\ms{A}}\mapsto \phi^{\ms{A}} +\frac{\pi}{2},
& &\phi^{\bar{\ms{A}}}\mapsto \phi^{\bar{\ms{A}}} -\frac{\pi}{2}.
\end{align} 
In order for the edge theory to be fully gapped one needs to add a second gapping term which needs to satisfy all the criteria mentioned in the main text. Additionally, 
 it needs to be invariant under the subgroup $\widetilde{\mathbb Z}_{2}$. A suitable gapping vector that can be added is
\begin{align}
\ell_{2}^{\ms{T}}=(2,2,-2,0),
\end{align}
which fully gaps the edge.

\section{Edge condensation between $\mathcal A$ and $\overline{\mathcal A}$}
In this appendix we describe the edge condensation procedure \cite{PhysRevB.79.045316, KONG2014436, Titus_Boson_2016} between time-reversal conjugate topological orders $\mathcal A$ and $ \bar{\mathcal A}$. By a folding trick \cite{Kapustin_2010}, the domain wall between $\mathcal A$ and $\bar{\mathcal A}$ is equivalent to the domain wall between $\mathcal A\times \mathcal A$ and the vacuum. More generally, folding reverses the orientation of a topological order. In euclidean topological quantum field theory (TQFT) different orientation reversing transformations such as time reversal and reflection may be treated on an equal footing as they are equivalent upto orientation-preserving transformations that act trivially on the theory. Since for both the higher-order topological insulator (HOTI) and higher-order topological superconductor (HOTSC), the surface topological order is chiral, the domain wall cannot be completely gapped. However as we will show, it is possible to condense a maximal subset of operators corresponding to bosonic and mutually local bulk anyons within $\mathcal A\times \mathcal A$, such that the edge/domain wall hosts a single chiral Dirac (resp. Majorana) mode when $\mathcal A$ corresponds to the surface topological order for HOTIs (resp. HOTSCs).  

Before describing the details of the condensation process for different choices of $\mathcal A$, we outline some generalities. There is a well established relationship between TQFTs in 2D and rational conformal field theories (CFTs) in 1D. An important fact that underlies this correspondence is that  line operators in the TQFTs and conformal blocks of the chiral algebra of the rational CFTs both separately give rise to an algebraic structure known as  a \emph{modular tensor category} (MTC) \cite{bakalov2001lectures, moore1990lectures, Kitaev_anyons}. This relationship begets a correspondence between bulk anyon condensation and edge condensation that we shall exploit. The condensation procedure may be briefly outlined as follows \cite{PhysRevB.79.045316, KONG2014436, Titus_Boson_2016}. First one identifies a set of objects to condense in the MTC that are bosonic (have integer topological spin) and mutually local (trivial $\ms{S}$-matrix). Let us denote this set of anyons as $\mathcal B$. Any two objects $a_1$ and $a_{2}$ that satisfy $a_1\in \mathcal B \times a_2$ are identified in the condensed theory, where the product `$\times$' corresponds to fusion in the MTC. If there exist such anyons $a_{1,2}$ with unequal topological spins, they get confined in the condensed theory. Finally, if $a$ appears in $ \mathcal B\times a$ with multiplicity $N$, then $a$ splits into $N+1$ objects in the condensed theory. Following this procedure one can obtain the objects within the condensed theory as well as all the additional data that goes into defining the condensed theory as an MTC.   

An equivalent algebraic recipe to study various condensations within anyon models was developed in Ref.~\onlinecite{Titus_Boson_2016}. We briefly describe it here for a condensation from topological order $\mathcal A\times \mathcal A$ to a topological order $\mathcal U$. Let the modular matrices corresponding to $\mathcal A\times \mathcal A$ and $\mathcal U$ be denoted by $(\ms{S},\ms{T})$ and $(\tilde{\ms{S}},\tilde{\ms{T}})$ respectively. Then one seeks a non-negative integer-valued symmetric square matrix $\ms{M}$ with $\ms{M}_{11}=1$ which commutes with $\ms{S}$ and $\ms{T}$. Given $\ms{M}$, there is a decomposition $\ms{M}=\ms{nn}^T$, such that $t\mapsto \sum_a \ms{n}_{a,t} a$ provides us with the \emph{lifting map} from $\mathcal{U}$ to $\mathcal A\times \mathcal A$, where $t\in \mathcal{U}$ and $a\in \mathcal{A}\times\mathcal{A}$. The topological data of $\mathcal{U}$ is constructed by solving for $\tilde{\ms{S}}$ and $\tilde{\ms{T}}$ using
\begin{equation}
\ms{S}\ms{n}=\ms{n}\tilde{\ms{S}},\quad \ms{T}\ms{n}=\ms{n}\tilde{\ms{T}},\quad \tilde{\ms{d}}_t=\frac{1}{q}\sum_a \ms{n}_{a,t} \ms{d}_a,
\label{solver_condensed}
\end{equation}
where $a\in \mathcal{A} \times \mathcal A$, $t\in \mathcal{U}$ and the normalization $q=\sum_a n_{a,1} d_a$ ensures that the the vacuum of the condensed theory has unit quantum dimension. 

\medskip Strictly speaking the topological orders appearing on the surface of both the HOTI and the HOTSC are not described by MTCs. This is because both these models contain a local fermion ($\psi_4$ in the $\mathcal T$-Pfaffian and $j=3$ in the $\ms{SO}(3)_6$ anyon model) and by definition each anyon/object that is not isomorphic to the vacuum within an MTC can be detected non-locally by at least one other object (or by itself) via a braiding operation. Since the aforementioned fermion is not detectable by braiding operations, in an MTC it should be identified with the vacuum; however, this is not possible because the vaccum is bosonic. Anyon models such as $\mathcal T$-Pfaffian or $\ms{SO}(3)_6$ are examples of super-MTCs \cite{Rubenstein_2016}: A super-MTC is a pre-modular tensor category with the property that there is a single (upto isomorphism) non-trivial object $\mathfrak f$, which is a local fermion, i.e., it has topological spin $-1$ and trivial braiding with all other anyons. For our purposes this distinction between MTCs and super-MTCs will not be very important as we will be able to extract the desired properties of the condensed theory using the  tools of anyon condensation for MTCs summarized above.

\section{Edge condensation on the surface of HOTI}
Let us consider the domain wall between two adjacent topological orders $\mathcal A$ and $\bar{\mathcal A}$ on the surface of a $\ms{C}_{2n}\mathcal T$-symmetric HOTI. As discussed in the main text, $\mathcal A$ is the anyon model corresponding to the $\mathcal T$-Pfaffian which contains a subset of the anyons in $\overline{\text{Ising}}\times \ms{U}(1)_8$. We denote the objects within the theory $\mathcal A\times \mathcal A$ by a subset of elements in the set 
\begin{align}
 \mathcal A\times \mathcal A \subset \left\{1_{j}^{\ms{A}},\psi_{j}^{\ms{A}},\sigma_{j}^{\ms{A}}\right\} \times \left\{1_{j}^{\bar{\ms{A}}},\psi_{j}^{\bar{\ms{A}}},\sigma_{j}^{\bar{\ms{A}}}\right\},
\label{eq:obj_AxA}
\end{align}
where $1,\psi$ and $\sigma$ label $\overline{\text{Ising}}$ anyons while $j\in \mathbb Z_{8}$ labels the $\ms{U}(1)_8$ anyons. The anyons in $\mathcal A$ are a subset of the 24 anyons in $\overline{\text{Ising}}\times \ms{U}(1)_8$ such that the $1$'s and $\psi$'s come with even $j$'s mod 8 while the $\sigma$'s come with odd $j$'s mod 8. The modular $\ms{S}$ and $\ms{T}$-matrices and fusion rules of the $\mathcal T$-Pfaffian model are inherited from the parent models which are well-known. For the $\overline{\text{Ising}}$-model, the fusion rules are
\begin{align}
1\times \sigma=1; \ 1\times \psi=1; \ \psi\times \psi=1;  \nonumber\\ \psi\times\sigma=\sigma;  \  \sigma\times \sigma =1+\psi,
\end{align}   
while the modular $\ms{S}$ and $\ms{T}$ matrices are
\begin{align}
\begin{split}
\ms{S}=&\,\begin{bmatrix}
\frac{1}{2}& \frac{\sqrt{2}}{2}& \frac{1}{2}\\
\frac{\sqrt{2}}{2} & 0 & -\frac{\sqrt{2}}{2}\\
\frac{1}{2}& -\frac{\sqrt{2}}{2}& \frac{1}{2}
\end{bmatrix},\quad
 \ms{T}=\begin{bmatrix}
1 & 0 &0\\
0 & e^{-\imagi\frac{\pi}{8}} & 0\\
0 &0 & -1
\end{bmatrix}.
\end{split}
\label{modular_Ising}
\end{align}
For the $\ms{U}(1)_8$ model, the fusion rules are $j_1\times j_2=j_1+j_2 \ \text{mod} \ 8$, while the modular $\ms{S}$ and $\ms{T}$-matrices are
\begin{align}
\ms{S}_{j_1j_2}=\exp\left\{\frac{2\pi ij_1j_2}{8}\right\}, \quad \ms{T}_{j_1j_2}=\exp\left\{\frac{\pi ij_1^2}{8}\right\}\delta_{j_1,j_2}. 
\label{modular_U(1)_8}
\end{align}
Within the $\mathcal T$-Pfaffian model, the $\ms{U}(1)_8$ sectors are charged such that the anyon $j$ carries a $j/4$ charge in units of $e$. Therefore in order for the fusion rules to be satisfied, there needs to be an underlying charge $2e$ condensate. Indeed, this is how the $\mathcal T$-Pfaffian model was originally motivated for the surface of the TI~\cite{Wang_2013}. 
 First, consider opening  a gap on the TI surface  by breaking charge $\ms{U}(1)$ symmetry via a charge $2e$-condensate induced by proximity to an s-wave superconductor. Since the goal is to construct a gapped symmetry-preserving surface, we must restore  $\ms{U}(1)$ symmetry  while leaving the surface gap closed. One route to this is to condense vortices of the superconductor. The underlying bulk topological response places constraints on the vortices that can be condensed: vortices with $2\pi n$ flux with $n$ odd always host a zero-energy $\mathcal{T}$-invariant Majorana Kramers doublet in the vortex core. Naively there does not seem to be any obstruction to  condensing the $4\pi$-flux vortex. However as argued in Ref.~\onlinecite{Wang_2013} this is precluded by the fact that the bulk $\mathbf{E}\cdot\mathbf{B}$ electromagnetic response of the TI leads to an effective Chern-Simons term for the surface theory that gives the $4\pi$ flux vortices  fermionic self-statistics. This can also be seen via a Berry phase computation in the $\ms{U}(1)$-broken the surface theory. Therefore the minimal condensable vortex is the bosonic $8\pi$ flux vortex, and upon condensation this leads to the $\ms{U}(1)_8$ topological order.

{\it A priori}, a domain wall between the 2D analogue $\mathcal{A}$ of the $\mathcal{T}$-Pfaffian and its time-reversal conjugate $\bar{\mathcal{A}}$ hosts a chiral conformal field theory such that each of the objects in Eq.~\eqref{eq:obj_AxA} represents a conformal character. These conformal characters are well-known quasiperiodic functions of the modular parameter on a torus~\cite{moore1990lectures,Ginsparg_CFT, francesco2012conformal} that reproduce Eq.~\eqref{modular_Ising} and Eq.~\eqref{modular_U(1)_8} upon modular transformations.
To begin, following Ref.~\onlinecite{Chen_2014}, we identify all the  bosonic anyons (i.e., those in the set $\mathcal B:=
\left\{1_{2}^{\ms{A}}\psi_{6}^{\bar{\ms{A}}}, 
\psi_{2}^{\ms{A}}1_{6}^{\bar{\ms{A}}},
\psi_{6}^{\ms{A}}1_{2}^{\bar{\ms{A}}},
1_{6}^{\ms{A}}\psi_{2}^{\bar{\ms{A}}}, 
1_{4}^{\ms{A}}1_{4}^{\bar{\ms{A}}},
\psi_{0}^{\ms{A}}\psi_{0}^{\bar{\ms{A}}},
\psi_{4}^{\ms{A}}\psi_{4}^{\bar{\ms{A}}}
\right\}$) with the vacuum. Thereafter, we can check that that only six sectors survive; each of these is a fusion orbit under the action of $ 1_{0}^{\ms{A}}1_{0}^{\bar{\ms{A}}}+ \mathcal B$ i.e each sector as a set is obtained by fusing a representative anyon with $ 1_{0}^{\ms{A}}1_{0}^{\bar{\ms{A}}}+ \mathcal B$, and can be labeled by a representative object from each orbit. In this notation we may denote the surviving orbits by 
\begin{align}
1_{0}^{\ms{A}}1_{0}^{\bar{\ms{A}}}, \psi_{0}^{\ms{A}}1_{0}^{\bar{\ms{A}}}, \psi_{4}^{\ms{A}}1_{0}^{\bar{\ms{A}}}, 1_{4}^{\ms{A}}1_{0}^{\bar{\ms{A}}}, \sigma_{1}^{\ms{A}}\sigma_{3}^{\bar{\ms{A}}},\sigma_{1}^{\ms{A}}\sigma_{7}^{\bar{\ms{A}}},
\end{align} 
which for brevity we shall shorten to
\begin{align}
1, \psi_{0}, \psi_{4}, 1_{4}, \sigma_{1}^{\ms{A}}\sigma_{3}^{\bar{\ms{A}}},\sigma_{1}^{\ms{A}}\sigma_{7}^{\bar{\ms{A}}}.
\label{eq:decon_sec}
\end{align}
Crucially, $\sigma_{1}^{\ms{A}}\sigma_{3}^{\bar{\ms{A}}}$ and $\sigma_{1}^{\ms{A}}\sigma_{7}^{\bar{\ms{A}}}$ split into two objects in the condensed theory, each of which is Abelian. More precisely $\sigma_{1}^{\ms{A}}\sigma_{3}^{\bar{\ms{A}}}$ splits into two Abelian anyons, each with charge $e$ and topological spin $-1$ while $\sigma_{1}^{\ms{A}}\sigma_{7}^{\bar{\ms{A}}}$ splits into two Abelian anyons each carrying charge $2e\sim 0$ and with topological spin $+1$. We denote the split sectors as
\begin{align}
\sigma^{\ms{A}}_{1}\sigma^{\bar{\ms{A}}}_{7}=&\; \alpha_1 + \alpha_2; \quad
\sigma^{\ms{A}}_{1}\sigma^{\bar{\ms{A}}}_{3}=\beta_1 + \beta_{2}.
\end{align}
The eight particles in the condensed theory  are listed, along with their charges, in Table~\ref{tab:cond_content}. The fusion rules of the surviving sectors are inherited from the parent theory $\mathcal A\times \mathcal A$. Notably the charge-neutral sectors form a fusion subalgebra (i.e. form a closed subset under fusion) given by
\begin{align}
\alpha_1 \times \alpha_1 =&\; 1,  \quad 
\alpha_2\times \alpha_2 = 1, \quad  
\alpha_1 \times \alpha_2= \psi_0,  \nonumber \\
\psi_0\times \psi_0=&\; 1, \quad 
\alpha_{1} \times \psi_0= \alpha_{2},  \quad 
\alpha_{2} \times \psi_0= \alpha_{1}. 
\end{align}
The $\ms{S}$-matrix of this theory can be obtained by using the Ribbon formula
\begin{align}
\ms{S}_{ij}=\frac{1}{\mathcal D}\sum_{k}\ms{N}_{ij}^{k}\frac{\theta_k}{\theta_i\theta_j}d_k.
\label{eq:ribbon}
\end{align}  
For example, it can be read off that the pairs of anyons $(\alpha_1,\alpha_2)$, $(\alpha_1,\psi_0)$ and $(\alpha_2,\psi_0)$ are mutual semions.  Therefore the fusion and braiding of the neutral anyons is equivalent to that of the toric code topological order. In fact upon compiling all the topological data, the condensed theory can be identified as a tensor product of the $\mathbb Z_{2}$-toric code with a local fermion $\left\{1,\mathfrak f\right\}$. The anyons may be labelled by elements in the set $\left\{1^0,e^0,m^0, f^0\right\}\times \left\{1,\mathfrak f\right\}$. The toric code sectors are charge neutral whereas the fermion $\mathfrak f$ carries electric charge 1. The identification with the anyons in Table~\ref{tab:cond_content} is
\begin{align}
\left\{1,\alpha_1,\alpha_2 , \psi_0\right\} \equiv&\;  \left\{1^{0},e^{0},m^{0},f^{0}\right\}, \nonumber \\
\left\{ \psi_4, \beta_2, \beta_1, 1_{4}\right\} \equiv&\;
\left\{1^{0},e^{0},m^{0},f^{0}\right\} \times \mathfrak f.
\label{eq:TPfaf_cond_sect}
\end{align}
\begin{table}[tb]
\begin{center}
 \begin{tabular}{ c |c | c | c | c | c | c | c | c } 
    $a\rightarrow$ & $1$ & $1_4$ & $\psi_0$ & $\psi_4$ & $\alpha_1$ & $\alpha_2$ & $\beta_1$ & $\beta_2$ \\ 
\hline
$e^{i\theta_a}$ & $+1$& $+1$ & $-1$ & $-1$  & $+1$ & $+1$ &  $-1$ & $-1$  \\ 
 \hline
$Q_a$ & 0& 1 & 0 & 1 & 0 & 0 & 1& 1 \\
\end{tabular}
\label{tab:cond_content}
\end{center}
\caption{Properties of anyons that survive after condensation of $\mathcal B$ in the anyon model \eqref{eq:obj_AxA}. }
\end{table}%
Finally we use the fact that chiral central charge is conserved in a condensation transition, therefore it can be read off that $\mathfrak f$ is a fermion with electric charge $e$ and chiral central charge $c_{-}=1$, i.e., a chiral Dirac fermion. 

\section{Edge condensation on the surface of HOTSC}
In this Section we consider the surface of a $\ms{C}_{2n}\mathcal T$ symmetric HOTSC. In particular, we focus on a single hinge between topological orders $\mathcal A$ and $\bar{\mathcal A}$ where $\mathcal A$ corresponds to the $\ms{SO}(3)_6$ anyon model \footnote{Sometimes also referred to as $\ms{PSU}(2)_6$.} which can be obtained from the $\ms{SU}(2)_6$ model by discarding all the half-integer representations. More precisely, the $\ms{SU}(2)_6$ model contains 7 anyons labelled as $j\in \{0,\frac{1}{2},1,\frac{3}{2},2,\frac{5}{2},3\}$.  The topological $\ms{S}$ and $\ms{T}$-matrices are
\begin{align}
\ms{S}_{j_1,j_2}=&\;\frac{1}{2}\sin\left[\frac{(2j_1+1)(2j_2+1)\pi}{8}\right] , \nonumber \\
\ms{T}_{j_1,j_2}=&\;e^{\frac{\imagi 2\pi j_1(j_1+1)}{8}}\delta_{j_1,j_2}.
\end{align}
The fusion rules  
\begin{align}
j_{1}\times j_{2}=\sum_{j=|j_1-j_2|}^{\text{min}\left\{j_1+j_2,k-j_1-j_2\right\}}j,
\end{align}
 are related to the $\ms{S}$-matrix via the Verlinde formula \cite{Verlinde_1988}.  The $j=0$ anyon is a boson and corresponds to the vacuum sector while the $j=3$ anyon is fermionic. In going from $\ms{SU}(2)_6$ to $\ms{SO}(3)_6$, all the anyons that braid non-trivially with  $j=3$ have been discarded. Therefore the $\ms{SO}(3)_6$ model contains four objects labelled $j=0,1,2,3$. With this, $\ms{SO}(3)_6$ contains a local fermion and is a super-MTC. The chiral central charge of $\ms{SO}(3)_6$ is $c=9/4$. The edge theory for the $\ms{SO}(3)_6$ topological order can be obtained as a quotient of the $\mathfrak{su}(2)_{6}$-Wess-Zumino-Witten model.

Having introduced the topological data and edge CFT corresponding to the $\ms{SO}(3)_6$ anyon model, we now turn to the anyon condensation within two copies of the model. We label sectors within the tensor product $\ms{SO}(3)_6\times \ms{SO}(3)_6$ by tuples $(jj')\in \{0,1,2,3\}\times \{0,1,2,3\}$. It is also straightforward to take a tensor product for the rest of the data defining the model. There are a total of 16 anyon sectors, four of which (i.e., $\left\{(00),(12),(21),(33)\right\}$) are bosonic and mutually local. Therefore these form a maximal set of condensable anyons. It can be shown explicitly that upon condensing all of the above bosons the fermions $\left\{(03),(30),(22),(11)\right\}$ are identified into a single sector while the remaining anyons are confined. It is illustrative to carry out this condensation procedure in two steps. First we condense the Abelian boson $(33)$. Upon doing so, pairs of anyons combine into single sectors. There are a total of eight sectors. Of these, $(00)\sim (33)$ and $(12)\sim (21)$ are bosons, $(03)\sim (30)$ and $(11)\sim (22)$ are fermions, $(10)\sim (23)$ and $(01)\sim (32)$ have topological spin $+i$, while $(02)\sim (31)$ and $(20)\sim(13)$ have topological spin $-i$. In the second condensation step the non-Abelian bosonic sector $(12)\sim (21)$ can be condensed, whereupon the two fermionic sectors are identified.  The final theory has a vacuum `$1$' and a transparent fermion `$\mathfrak f$' with a  lifting map $\ms{n}^{T}$ given by
\begin{align}
1\mapsto&\;  (00) + (33) + (12) + (21), \nonumber \\ 
f\mapsto&\; (03) + (30) + (11) + (22). 
\end{align}
The chiral central charge $c_{-}=9/2$ can be read off from the pre-condensed theory and  is twice the chiral central charge of the $\ms{SO}(3)_6$ anyon model. The chiral central charge of the HOTSC hinge is only stable modulo integers as one can always add/ remove two chiral majorana hinge modes (i.e $c_{-}=1$) by pasting $\ms{p}\pm i\ms{p}$ phases on adjacent surfaces in a $\ms{C}_{2n}\mathcal T$-symmetric manner. Consequently a single chiral majorana hinge mode that is stable in the weakly interacting regime can be unhinged by the chiral fermion $\mathfrak f$ without breaking $\ms{C}_{2n}\mathcal T$-symmetry.

\section{$\ms{C}_{2n}\mathcal T$ symmetric HOSPT}

As we noted in our conclusions, our approach can be readily adapted to study various surface terminations of interacting bosonic HOSPTs with $\ms{C}_{2n}\mathcal T$ symmetry. As an illustration we present one such construction here. We can construct $\ms{C}_{2n}\mathcal T$ HOSPT by starting from a 3D $\mathcal T$-symmetric bosonic symmetry-protected state (SPT). We focus on a particularly simple example: the so-called ``bosonic topological superconductor'', protected by time-reversal symmetry $\mathcal{T}$ (with $\mathcal{T}^2=1$ as appropriate to a bosonic system.) The term superconductor is appropriate because the system does {\it not} need to satisfy $U(1)$ charge conservation. This phase was conjectured via field-theoretic arguments in Ref.~\onlinecite{Vish_2013} and given an explicit lattice construction via a Walker-Wang model in Ref.~\onlinecite{Burnell_2014}.  Despite its simplicity, this phase lies outside the ``group cohomology'' classification of bosonic SPTs. Instead, it motivates a distinct perspective on SPTs based on the mathematical framework of cobordism theory~\cite{2014arXiv1403.1467K,2014arXiv1404.6659K}.

 When placed on a manifold with boundaries, the bosonic TSC hosts a gapless surface, whose properties are best characterized for our present purposes by the fact that it exhibits a half-quantized bosonic thermal Hall effect (upon breaking $\mathcal{T}$). It is useful to clarify this statement further. It is known that the thermal Hall conductance of any purely bosonic 2D system without fractionalized bulk excitations is forced to be quantized as $\frac{\kappa_{xy}}{T} = 8n$ with  $n\in \mathbb{Z}$ and in units of $\pi^2 k_B^2/3h$, where  $n=1$ case is realized by the Kitaev $\ms{E}_8$ state~\cite{Kitaev_anyons}, an invertible topological order~\footnote{An invertible topological order is one with no fractionalized bulk excitations and possibly a non-zero chiral central charge. Examples include the $\ms{E}_8$ state of bosons, the integer quantum Hall effect, and the $\ms{p} + i\ms{p}$ superconductor.} with  $8$ chiral bosons at the edge. If we break $\mathcal{T}$ symmetry on the bosonic TSC surface, a domain wall between opposite $\mathcal{T}$-breaking regions  necessarily traps a set of chiral bosonic modes with chiral central charge $c_{-}=8$. Since the two domains are linked by $\mathcal{T}$-symmetry, they can each be assigned `half' the $\ms{E}_8$ edge, and hence a surface with a single $\mathcal{T}$-breaking domain can be viewed as having a `half-quantized' thermal Hall effect of bosons.
 
 To build the HOTSC we break the full combination of time reversal and rotation symmetry about a certain axis to a subgroup $\ms{C}_{2n}\mathcal T$.  This symmetry pins the $\mathcal{T}$-breaking domain wall to the hinges, which thus carry $\ms{E}_8$-chiral modes in a $\ms{C}_{2n}\mathcal T$ symmetric pattern. Since the simplest non-fractionalized 2D  state of bosons is $\mathcal{T}$-breaking and has $c_-=8$, any non-fractionalized surface termination that preserves $\ms{C}_{2n}\mathcal T$ can only change the hinge central charge in units of $\Delta c_- =16n$, so that without fractionalization the hinge mode is globally stable as long as  $\ms{C}_{2n}\mathcal T$ is preserved. We have
 thus constructed a $\ms{C}_{2n}\mathcal T$  `bosonic HOTSC'. Note that a very similar similar construction of a bosonic point-group SPT protected by rotation/mirror symmetries was provided in Ref.~\cite{PhysRevX.7.011020}.

Next, we ask what $\ms{C}_{2n}\mathcal T$-symmetric surface topological order can absorb the $\ms{E}_8$ hinge modes. We return to the first-order case, and observe that 
its symmetry-preserving surface topological order (STO) $\mathcal{A}_{\mathcal{T}}$ is the ``three-fermion $\mathbb Z_{2}$ toric code'' topological order, which has the correct anomaly \cite{Vish_2013, Burnell_2014} to match the bulk response. Following our successful strategy in the HOTI/HOTSC cases, we propose placing  $\mathcal{A}_{\mathcal{T}}$ on the top/bottom surfaces, and pattern alternating sides with $\mathcal{A}$ and $\bar{\mathcal{A}}$ that we take to be the 2D $\mathcal{T}$-breaking analogues of $\mathcal{A}_{\mathcal{T}}$. We now briefly summarize the properties of these `three fermion toric code'  topological orders. The bulk anyon content is common to all three theories  $\mathcal{A}_{\mathcal{T}}$, $\mathcal{A}$ and $\bar{\mathcal{A}}$ and is given by the
 $\ms{SO}(8)_1$ topological order~\cite{Vish_2013, Burnell_2014}, described by an Abelian Chern-Simons theory with $K$-matrix
\begin{align}\label{eq:SO8K}
K_{\ms{SO}_8}=\begin{bmatrix}
2 & -1 &-1& -1\\
-1 & 2 & 0 & 0\\
-1 & 0 & 2 & 0\\
-1 & 0 & 0 & 2\\
\end{bmatrix}.
\end{align}   
 The theory has four anyons, whose fusion rules are analogous to the $\mathbb Z_{2}$ toric code with the exception that the $e$ and $m$ particles are fermionic --- hence its name. We label the particle types as $1^{\ms{A}},f^{\ms{A}}_{i}$ where $i=1,2,3$ (and all the $f_{i}^{\ms{A}}$ are fermions). Time reversal does not permute the anyons, and squares to identity on all three anyons. While  the time-reversal invariant theory $\mathcal{A}_{\mathcal{T}}$ can only realized on the surface of a 3D bosonic TSC, $\mathcal{A}$ and $\bar{\mathcal{A}}$ can be realized in 2D, in which case they have a chiral edge to vacuum described by the $K$-matrix (\ref{eq:SO8K}) with chiral central charge $c_-=4$ --- precisely one half of that bound to the hinge of our 2D bosonic HOTSC.
 
 We now demonstrate that the side hinges are indeed gapped by this construction. To analyze the hinge, it suffices to simply consider the folded theory $\mathcal A\times \mathcal A$ which has sixteen anyons labelled by elements in the set $\left\{1^{\ms{A}},f^{\ms{A}}_1,f^{\ms{A}}_2,f^{\ms{A}}_3\right\}\times \left\{1^{\bar{\ms{A}}},f^{\bar{\ms{A}}}_1,f^{\bar{\ms{A}}}_2,f^{\bar{\ms{A}}}_3\right\}$. The hinge between $\mathcal A$ and $\bar{\mathcal A}$ hosts a chiral Luttinger liquid with $K=K_{\ms{SO}_8}\oplus K_{\ms{SO}_8}$. This hinge can be reduced to the edge of the $\ms{E}8$ state upon condensing $\mathcal B=\left\{f_{1}^{\ms{A}}f_{1}^{\bar{A}}, f_{2}^{\ms{A}}f_{2}^{\bar{A}}, f_{3}^{\ms{A}}f_{3}^{\bar{A}}\right\}$.  Precisely in analogy to the construction of gapped surfaces for HOTI and HOTSC, one obtains a completely gapped surface termination for the bosonic HOTSC constructed above.
 
 For completeness, we provide a complementary analysis using the conceptually simpler but more tedious approach based on chiral Luttinger liquids. The hinge contains degrees of freedom contributed by the topological order $\mathcal A$ and $\bar{\mathcal A }$ as well as the `$\ms{E}_8$-hinge mode' contributed by the bulk HOSPT. Altogether the hinge is described by a chiral Luttinger liquid with matrix $K=K_{\ms{SO}(8)}\oplus K_{\ms{SO}(8)} \oplus K_{\overline{\ms{E}}_8}$ where 
 \begin{align}
 K_{\overline{\ms{E}}_8}=
 \begin{bmatrix}
-2 & 1 & 0 & 0 & 0 & 0 & 0 & 0\\
1  &-2 & 1 & 0 & 0 & 0 & 0 & 0\\
0  & 1 &-2 & 1 & 0 & 0 & 0 & 0\\
0  & 0 & 1 & -2 & 1 & 0 & 0 & 0\\
0  & 0 & 0 &  1 & -2 & 1 & 0 & 1\\
0  & 0 & 0 & 0 & 1 & -2 & 1 & 0\\
0  & 0 & 0 & 0 & 0 & 1 & -2 & 0\\
0  & 0 & 0 & 0 & 1 & 0 & 0 & -2\\
\end{bmatrix}.
 \end{align}
As before this theory may be gapped by adding terms of the form $\Delta \mathcal L=\sum_{i=1}^{8}\lambda_{i}\cos\left[\ell_{i}^{\ms{T}}\Phi + \alpha_i\right]$ where $\Phi$ is a sixteen component vector of compact bosons in the natural basis of $K$. A possible choice of $\left\{\ell_i\right\}$, the set of gapping vectors satisfying constraints $\ell_{i}^{\ms{T}}K^{-1}\ell_j=0$ and $\ell_{i}\in K\mathbb Z^{16}$ for all $i,j$ are
\begin{align}
\ell_1^{\ms{T}}=&\;  \left( \begin{array}{cccccccccccccccc}
2 & -1 & -1 & -1 &
0 & 0 & 0 & 0 &
1 & -2 & 1 & 0 &
0 & 0 & 0 & 0 
 \end{array} \right) \nonumber \\
 \ell_2^{\ms{T}}=&\;  \left( \begin{array}{cccccccccccccccc}
-1 & 2 & 0 & 0 &
0 & 0 & 0 & 0 &
-2 & 1 & 0 & 0 &
0 & 0 & 0 & 0 
 \end{array} \right) \nonumber \\
 \ell_3^{\ms{T}}=&\;  \left( \begin{array}{cccccccccccccccc}
0 & 1 & 1 & -1 &
-1 & 0 & 2 & 0 &
-1 & 0 & 1 & -2 &
0 & 1 & 0 & 1 
 \end{array} \right) \nonumber \\
 \ell_4^{\ms{T}}=&\;  \left( \begin{array}{cccccccccccccccc}
-1 & 0 & 0 & 2 &
0 & 0 & 0 & 0 &
0 & 1 & -2 & 1 &
0 & 0 & 0 & 0 
 \end{array} \right) \nonumber \\
 \ell_5^{\ms{T}}=&\;  \left( \begin{array}{cccccccccccccccc}
0 & 0 & 0 & 0 &
2 & -1 & -1 & -1 &
0 & 0 & 0 & 0 &
1 & -2 & 1 & 0 
 \end{array} \right) \nonumber \\
 \ell_6^{\ms{T}}=&\;  \left( \begin{array}{cccccccccccccccc}
0 & 0 & 0 & 0 &
-1 & 2 & 0 & 0 &
0 & 0 & 0 & 1 &
-2 & 1 & 0 & 1 
 \end{array} \right) \nonumber \\
 \ell_7^{\ms{T}}=&\;  \left( \begin{array}{cccccccccccccccc}
-2 & 0 & 2 & 2 &
-2 & 2 & 2 & 0 &
0 & 2 & -2 & 0 &
-2 & 2 & 0 & 2 
 \end{array} \right) \nonumber \\
  \ell_8^{\ms{T}}=&\;  \left( \begin{array}{cccccccccccccccc}
0 & 0 & 0 & 0 &
-1 & 0 & 0 & 2 &
0 & 0 & 0 & 0 &
0 & 1 & -2 & 0 
 \end{array} \right) .
\end{align}
Note that there is no $\ms{U}(1)$ symmetry charge to consider here so we only need to consider the compatibility of the different gapping vectors.

\end{widetext}
\end{document}